\documentclass[final]{siamltex}

\usepackage{euscript,amsmath,amssymb,amsfonts,graphicx}
\allowdisplaybreaks[1]

\usepackage{chngcntr}
\counterwithout{figure}{section}
\counterwithout{equation}{section}

\usepackage{epstopdf}

\newcommand{\e}{{\rm e}}
\renewcommand{\d}{{\rm d}}
\newcommand{\pd}{\partial}

\newcommand{\mc}{\mathcal }
\newcommand{\ve}{\varepsilon}

\title{Short term synaptic depression improves information transfer in perceptual multistability}

\author{Zachary P. Kilpatrick} 

\begin{document}

\maketitle

\begin{abstract} 
Competitive neural networks are often used to model the dynamics of perceptual bistability. Switching between percepts can occur through fluctuations and/or a slow adaptive process. Here, we analyze switching statistics in competitive networks with short term synaptic depression and noise. We start by analyzing a ring model that yields spatially structured solutions and complement this with a study of a space-free network whose populations are coupled with mutual inhibition. Dominance times arising from depression driven switching can be approximated using a separation of timescales in the ring and space-free model. For purely noise-driven switching, we use energy arguments to justify how dominance times are exponentially related to input strength. We also show that a combination of depression and noise generates realistic distributions of dominance times. Unimodal functions of dominance times are more easily differentiated from one another using Bayesian sampling, suggesting synaptic depression induced switching transfers more information about stimuli than noise-driven switching. Finally, we analyze a competitive network model of perceptual tristability, showing depression generates a memory of previous percepts based on the ordering of percepts. 
\end{abstract}

\begin{keywords}
binocular rivalry, neural field, ring model, bump attractor, short term depression
\end{keywords}

\section{Introduction}
\label{intro}
Ambiguous sensory stimuli with two interpretations can produce perceptual rivalry \cite{blake02}. For example, two orthogonal gratings presented to either eye lead to perception switching between one grating and then the next repetitively, a common paradigm known as binocular rivalry \cite{leopold96}. Perceptual rivalry can also be triggered by a single stimulus with two interpretations, like the Necker cube \cite{orbach63}. One notable feature of the switching process in perceptual rivalry is its stochasticity -- a histogram of the dominance times of each percept spreads across a broad range \cite{fox67}. Senses other than vision also exhibit perceptual rivalry. When two different odorants are presented to the two nostrils, a similar phenomenon occurs with olfaction, termed ``binaral" rivalry \cite{zhou09}. Similar experiences have been evoked in the auditory \cite{deutsch74,pressnitzer06} and tactile \cite{carter08} system.

Multiple principles govern the relationship between the strength of percepts and the mean switching statistics in perceptual rivalry \cite{levelt65}. ``Levelt's propositions" relate stimulus contrast to the {\em mean dominance times}: (i) increasing the contrast of one stimulus increases the proportion of time that stimulus is dominant; (ii) increasing the contrast of one stimulus does not affect its average dominance time; (iii) increasing contrast of one stimulus increases the rivalry alternation rate; and (iv) increasing the contrast of both stimuli increases the rivalry alternation rate. There are also relationships between the properties of the input and the stochastic variation in the dominance times \cite{brascamp06}. For instance, the distribution of dominance times is well fit by a gamma distribution \cite{fox67,lehky95,vanee09}. The fact that dominance times are not exponentially distributed suggests some background slow adaptive process plays a role in providing a nonzero peak in the dominance histograms \cite{shpiro09}. Two commonly proposed mechanisms for this adaptation are spike frequency adaptation and short term synaptic depression \cite{laing02,wilson03,shpiro07}. An even stronger case can be made of the existence of adaptation in perceptual processing networks by examining results of experiments on {\em perceptual tristability} \cite{hupe10}. Here, perception alternates between three possible choices and subsequent switches are determined by the previous switch \cite{naber10}. This memory suggests switches in perceptual multistability are not purely noise-driven \cite{moreno07}.

Most theoretical models of perceptual rivalry employ two pools of neurons, each selective to one percept, coupled to one another by mutual inhibition \cite{matsuoka84,laing02,shpiro07,seely11}. With no other mechanisms at work, such architectures lead to {\em winner-take-all} states, where one pool of neurons inhibits the other indefinitely \cite{wang92}. However, switches between the dominance of one pool and the other can be initiated with the inclusion of fluctuations \cite{moreno07} or an adaptive process \cite{laing02,shpiro07}. Combining the two effects leads to dominance times that are distributed according to the gamma distribution \cite{laing02,shpiro09,vanee09}. Slow adaptation and noise thus serve as agents for the sampling of the stimulus through network activity. A mutual inhibitory network would otherwise remain in the {\em winner-take-all} state indefinitely.

In light of these observations, we wish to consider the role adaptive mechanisms play in properly sampling ambiguous stimuli in the context of a mutual inhibitory network. Purely fluctuation driven switching would provide a noisy sample of the two percepts, but pure adaptation would provide an extremely reliable sampling of percept contrast \cite{shpiro09}. Thus, as the level of adaptation is increased and noise is decreased, one would expect that the ability of mutual inhibitory networks to encode information about ambiguous stimuli is vastly improved. A major point is that it is also vastly improved over networks without any adaptation at all. We focus specifically on short term synaptic depression \cite{abbott97,tsodyks98}.

Using parametrized models, we will explore how synaptic depression improves the ability of a network to extract stimulus contrasts. First, we will be concerned with how much information can be determined about the contrast of each of the two percepts of an ambiguous stimulus. In the case of a {\em winner-take-all} solution, only information about a single percept could be known, since the pool of neurons encoding the other percept would be quiescent. We will study this problem using an anatomically motivated neural field model of an orientation column with synaptic depression \cite{york09,kilpatrick10b}, given by (\ref{ringmod}). We find that increasing the strength of synaptic depression from zero leads to a bifurcation whereby rivalrous oscillations onset. When rivalrous switching occurs through a combination of depression and noise, we show stronger depression improves the transfer of information using simple Bayesian inference \cite{kersten04}. We also analyze a competitive network model with depression and noise (\ref{fulcompnet}) to help study the combined effects of noise and depression on perceptual switching. In particular, we will show that the presence of synaptic depression increases the information relayed by the output of the network. Finally, we will show trimodal stimuli to a neural field model with synaptic depression can generate oscillations where each mode spends time in dominance. To deeply analyze the relative contributions of noise and depression to this switching process, we study a reduced model. This reveals depression generates a history dependence in switching that would not arise in the network with purely noise-driven switching.

\section{Materials and Methods}

\subsection{Ring model with synaptic depression}

As a starting point, we will consider a model for processing the orientation of visual stimuli \cite{benyishai95,bressloff02} which also includes short term synaptic depression \cite{york09,kilpatrick10b}. Since GABAergic inhibition is much faster than AMPA-mediated excitation \cite{kawaguchi97}, we make the assumption that inhibition is slaved to excitation as in \cite{amari77}. Reduction this disynaptic pathway and assuming depression acts on on excitation \cite{tsodyks98}, we then have the model \cite{york09,kilpatrick10}
\begin{subequations}  \label{ringmod}
\begin{align}
\tau_m \frac{\pd u(x,t)}{\pd t} &= - u(x,t) + \int_{- \pi / 2}^{\pi / 2} w( x - y) q(y,t) f(u(y,t)) \d y + I(x) + \xi (x,t), \\
\tau \frac{\pd q(x,t)}{\pd t} &= 1 - q(x,t) - \beta q(x,t) f(u(x,t)).
\end{align} 
\end{subequations}
Here $u(x,t)$ measures the synaptic input to the neural population with stimulus preference $x$ at time $t$, evolving on the timescale $\tau_m$. Firing rates are given by taking the gain function $f(u)$ of the synaptic input, which we usually proscribe to be \cite{wilson73}
\begin{align}
f(u) = \frac{1}{1+ \e^{- \gamma (u- \kappa)}}, \label{sig}
\end{align}
and often take the high gain limit $\gamma \to \infty$ for analytical convenience, so \cite{amari77,kilpatrick10}
\begin{align}
f(u) = H(u- \kappa) = \left\{ \begin{array}{ll} 0 & : u < \kappa, \\ 1 & : u \geq \kappa. \end{array} \right.  \label{H}
\end{align}
External input, representing flow from upstream in the visual system is prescribed by the time-independent function $I(x)$ \cite{benyishai95,bressloff02}. For the majority of our study of (\ref{ringmod}), we employ the bimodal stimulus
\begin{align}
I(x) = -I_0 \cos (4x) + I_a \sin (2x),  \label{bimod}
\end{align}
representing stimuli at the two orthogonal angles $- \pi /4$ and $\pi / 4$ and $I_0$ controls the mean of each peak and $I_a$ controls the level of asymmetry between the peaks. Effects of noise are described by the stochastic processe $\langle \xi (x,t) \rangle$ with $\langle \xi (x,t) \rangle = 0$ and $\langle \xi (x,t) \xi (y,s) \rangle = C(x-y) \delta (t-s)$. For simplicity, we assume the spatial correlations have a cosine profile $C(x) = \pi \cos (x)$. Synaptic interactions are described by the integral term, so $w(x-y)$ describes the strength (amplitude of $w$) and net polarity (sign of $w$) of synaptic interactions from neurons with stimulus preference $y$ to those with preference $x$. Following previous studies, we presume the modulation of the synaptic strength is given by the cosine
\begin{align}
w(x-y) = \cos (2(x-y)),   \label{cos}
\end{align}
so neurons with similar orientation preference excite one another and those with dissimilar orientation preference disynaptically inhibit one another \cite{benyishai95,ferster00}. The factor $q(x,t)$ measures of the fraction of available presynaptic resources, which are depleted at a rate $\beta f$ \cite{abbott97,tsodyks98}, and are recovered on a timescale specified by the time constant $\tau$ \cite{chance98}.

By setting $\tau_m=1$, we can assume time evolves on units of the excitatory synaptic time constant, which we presume to be roughly 10ms \cite{hausser97}. Experimental observations have shown synaptic resources specified $q$ are recovered on a timescale of 200-800ms \cite{abbott97,tsodyks97}, so we require $\tau$ is between 20 and 80, usually setting it to be $\tau = 50$. Our parameter $\beta$ can then be varied independently to adjust the effective depletion rate of synaptic depression.

\subsection{Idealized competitive neural network}
We also study space-free competitive neural networks with synaptic depression \cite{shpiro07}. In this way, we can make more progress analyzing switching behavior. As a general model of networks connected by mutual inhibition, we consider the system \cite{laing02,moreno07,shpiro07}
\begin{subequations}  \label{fulcompnet}
\begin{align}
\dot{u}_R(t) &= - u_1(t) + f(I_R - q_L(t) u_L(t) ) + \xi_1(t), \\
\dot{u}_L (t) &= - u_2(t) + f(I_L - q_R(t) u_R(t)) + \xi_2 (t), \\
\tau \dot{q}_R(t) & = 1 - q_R(t) - \beta u_R(t) q_R(t), \\
\tau \dot{q}_L(t) &= 1 - q_L (t) - \beta u_L (t) q_L (t), 
\end{align}
\end{subequations}
where $u_j$ represents the firing rate of the $j=1,2$ population. The strength of recurrent synaptic excitation within a population is specified by the parameter $\alpha$, whereas the strength of cross-inhibition between populations is specified by $\beta$. Fluctuations are introduced into population $j$ with the independent white noise processes $\xi_j$ with $\langle x_j (t) \rangle = 0$ and $\langle \xi_j(t) \xi_j(s) \rangle = \ve \delta (t - s)$.

\subsection{Numerical simulation of stochastic differential equations}

The spatially extended model (\ref{ringmod}) was simulated using an Euler-Maruyama method with a timestep {\tt dt} $=10^{-4}$, using Riemann integration on the convolution term with {\tt N} $= 2000$ spatial grid points. The space clamped competitive network (\ref{fulcompnet}) was also simulated using Euler-Maruyama with a timestep {\tt dt} $=10^{-6}$. To generate histograms of dominance times, we simulated systems for 20000s ($2 \times 10^6$ time units).

\subsection{Fitting dominance time distributions}

To generate the theoretical curves presented for exponentially distributed dominance times, we simply take the mean of dominance times and use it as the scaling in the exponential (\ref{switchexp}). For those densities that we presume are gamma distributed, we solve a linear regression problem. Specifically, we look for the constants $c_1$, $c_2$, and $c_3$ of
\begin{align}
f(T) = \e^{c_1} T^{c_2} \e^{- c_3 T}  \label{altgam}
\end{align}
an alternate form of (\ref{gamdist}). Upon taking the logarithm of (\ref{altgam}), we have the linear sum
\begin{align}
\ln f(T) = c_1 + c_2 \ln T - c_3 T. \label{sumgam}
\end{align}
Then, we select three values of the numerically generated distribution $p^n(T^n)$ along with its associated dominance times: $(T^n_1,p^n_1); (T^n_2,p^n_2); (T^n_3,p^n_3)$ where $p^n_j = p^n(T^n_j)$. We always choose $T^n_2 = \arg \max_T p^n (T)$ as well as $T^n_1 = T^n_2/2$ and $T^n_3 = 3 T^n_2/2$. It is then straightforward to solve the linear system
\begin{align*}
\left( \begin{array}{ccc} 1 & \ln T^n_1 & - T^n_1 \\ 1 & \ln T^n_2 &  - T^n_2 \\ 1 & \ln T^n_3 & - T^n_3  \end{array} \right) \left( \begin{array}{c} c_1 \\ c_2 \\ c_3 \end{array} \right) = \left( \begin{array}{c} \ln p^n_1 \\ \ln p^n_2 \\ \ln p^n_3  \end{array} \right)
\end{align*}
for the associated constants using the {\small \bf \textbackslash } command in {\tt MATLAB}.

\section{Results}

We now discuss several results that reveal the importance of synaptic depression in transferring information about stimuli to competitive networks. This is initially shown by analyzing the ring model with depression (\ref{ringmod}). However, to carry out detailed analysis on stochastic switching a competitive network with depression, we must reduce (\ref{ringmod}) as well as analyzing an analogous model without space (\ref{fulcompnet}).

\subsection{Deterministic switching in the ring model}

To start we will consider the ring model with depression (\ref{ringmod}) in the absence of noise. We then have the deterministic system \cite{york09,kilpatrick10,kilpatrick10c}
\begin{subequations}  \label{ringdet}
\begin{align}
u_t (x,t) &= - u(x,t) + \int_{- \pi/2}^{\pi/2} w(x-y) q(y,t) f(u(y,t)) \d y + I(x), \\
\tau q_t (x,t) &= 1 - q(x,t) - \beta q(x,t) f(u(x,t)).
\end{align}
\end{subequations}
In previous work, versions of (\ref{ringdet}) have been analyzed to explore how synaptic depression can generate traveling pulses \cite{york09,kilpatrick10}, self-sustained oscillations \cite{kilpatrick10}, and spiral waves in two-dimensions \cite{kilpatrick10c}. Here, we will extend previous work that explored input-driven oscillations in two-layer networks like (\ref{ringdet}) that possessed many statistics matching binocular rivalry \cite{kilpatrick10b}. We will think of (\ref{ringdet}) as a model of {\em monocular rivalry}, since oscillations can be due to competition between representations in a single orientation column \cite{benyishai95}. Competition between ocular dominance columns \cite{kilpatrick10b} is not necessary for our theory. For the purpose of exposition, we will employ specific forms for the functions of (\ref{ringdet}): cosine weight (\ref{cos}); a Heaviside firing rate function (\ref{H}); and a bimodal input (\ref{bimod}).

\begin{figure}
\begin{center} \includegraphics[width=13cm]{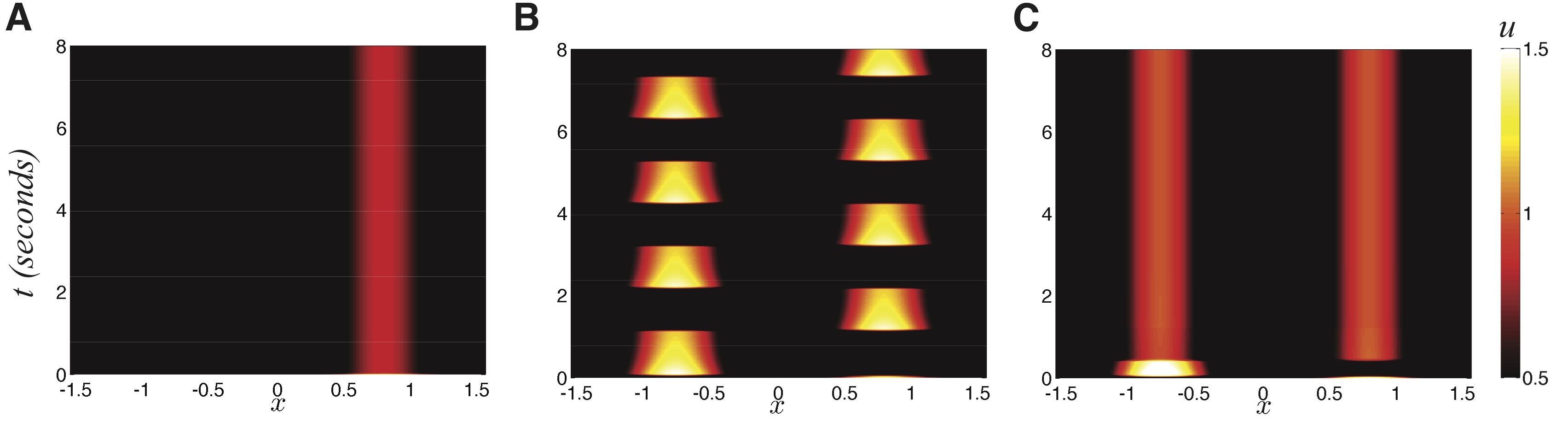} \end{center}
\caption{Three possible active states of the noise-free stimulus driven ring model with depression (\ref{ringdet}). {\bf (A)} Winner take all ($I_0 = 0.6$); {\bf (B)} Rivalrous oscillations ($I_0 = 0.84$); {\bf (C)} Fusion ($I_0 = 1$). Other parameters are $\theta = 0.5$, $\beta =1$, and $\tau = 50$.}
\label{rdstates}
\end{figure}

{\bf Winner take all state.} We start by looking for winner-take-all solutions to the deterministic system (\ref{ringdet}), such as that shown in Fig. \ref{rdstates}{\bf (A)}. These states consist of a single activity bump arising in the network, representing only one of the two percepts contained in the bimodal stimulus (\ref{bimod}). These are stationary in time, so $u_t = q_t = 0$, implying that $u=U(x)$ and $q=Q(x)$. Also, they are single bump solutions, so there will be a single region in space that is superthreshold ($U(x)> \kappa$). We assume the right stimulus is represented by a bump, although we can derive analogous results when the left stimulus is represented. We then have a steady state solution determined by
\begin{align}
U(x) & = \int_{\pi/4 - a}^{\pi/4 + a} \cos (2(x-y)) Q(y) \d y - I_0 \cos (4x) + I_a \sin (2x),  \label{wtau} \\
Q(x) &= \left[ 1 + \beta H(U(x)- \kappa ) \right]^{-1}, \label{wtaq}
\end{align}
since $U(x)> \kappa$ when $x \in ( \pi/4 -a, \pi/4 + a)$, so that by plugging (\ref{wtaq}) into (\ref{wtau}) and using the trigonometric identity $\cos (2(x-y)) = \cos (2x) \cos (2y) + \sin (2x) \sin (2y)$ we have
\begin{align}
U(x) = A \cos (2x) + B \sin (2x) - I_0 \cos (4x) + I_a \sin (2x),
\end{align}
where the multiplicative constants $A,B$ can be computed
\begin{align*}
A = \frac{1}{1 + \beta} \int_{\pi/4 - a}^{\pi/4 + a} \cos (2x) \d x = 0, \ \ \ \ \ \ B = \frac{1}{1+ \beta} \int_{\pi/4 - a}^{\pi/4 + a} \sin (2x) \d x = \frac{\sin (2a)}{1 + \beta}.
\end{align*}
Therefore, by simplifying the threshold condition, $U( \pi / 4 \pm a) = \kappa$, we have
\begin{align}
U( \pi /4 \pm a ) &= \frac{\sin (4a)}{2(1+ \beta)} + I_0 \cos (4a) + I_a \cos (2a) = \kappa. \label{wtasc}
\end{align}
The implicit equation (\ref{wtasc}) can be solved numerically using root finding algorithms. For symmetric inputs ($I_a = 0$), we can solve (\ref{wtasc}) explicitly
\begin{align}
a = \frac{1}{2} \tan^{-1} \left[ \frac{1 \pm \sqrt{1 + 4 (1+ \beta)^2 (I_0^2 - \kappa^2) }}{2(1+ \beta ) ( I_0 + \kappa )} \right]. \label{wtawid}
\end{align}
Thus, we can fully characterize winner-take-all solutions
\begin{align}
U(x) = \frac{\sin (2a)}{1+ \beta} \sin (2x) - I_0 \cos (4x) + I_a \sin (2x).  \label{wtasol}
\end{align}
The advantage of having this solution is that we can relate the parameters of the model to the existence of the winner-take-all state, where we would expect to only see single bump solutions. To do so, we need to look at a second condition that must be satisfied, $U(x)< \kappa$ for all $x \notin ( \pi / 4 - a, \pi / 4 + a)$. Since the function (\ref{wtasol}) is bimodal across $(-\pi/2 , \pi / 2)$, we check the other possible local maximum at $x = - \pi /4$ as
\begin{align}
U( \pi /4 ) = I_0-I_a - \frac{\sin (2a)}{1+ \beta}  < \kappa. \label{wtasub}
\end{align}
At the point in parameter space where the (\ref{wtasub}) is violated, a bifurcation occurs, so the winner-take-all state ceases to exist. This surface in parameter space is given by the equation
\begin{align}
I_0 = \kappa + I_a + \frac{\sin (2a)}{1+ \beta},  \label{wtabif}
\end{align}
along with the explicit formula for the bump half-width (\ref{wtawid}). Beyond the bifurcation boundary (\ref{wtabif}), one of two behaviors can occur. Either there is a symmetric two-bump solution that exists, the fusion state \cite{wolfe86,blake89,shpiro07}, or rivalrous oscillations \cite{levelt65,blake02}.

\begin{figure}
\begin{center} \includegraphics[width=13cm]{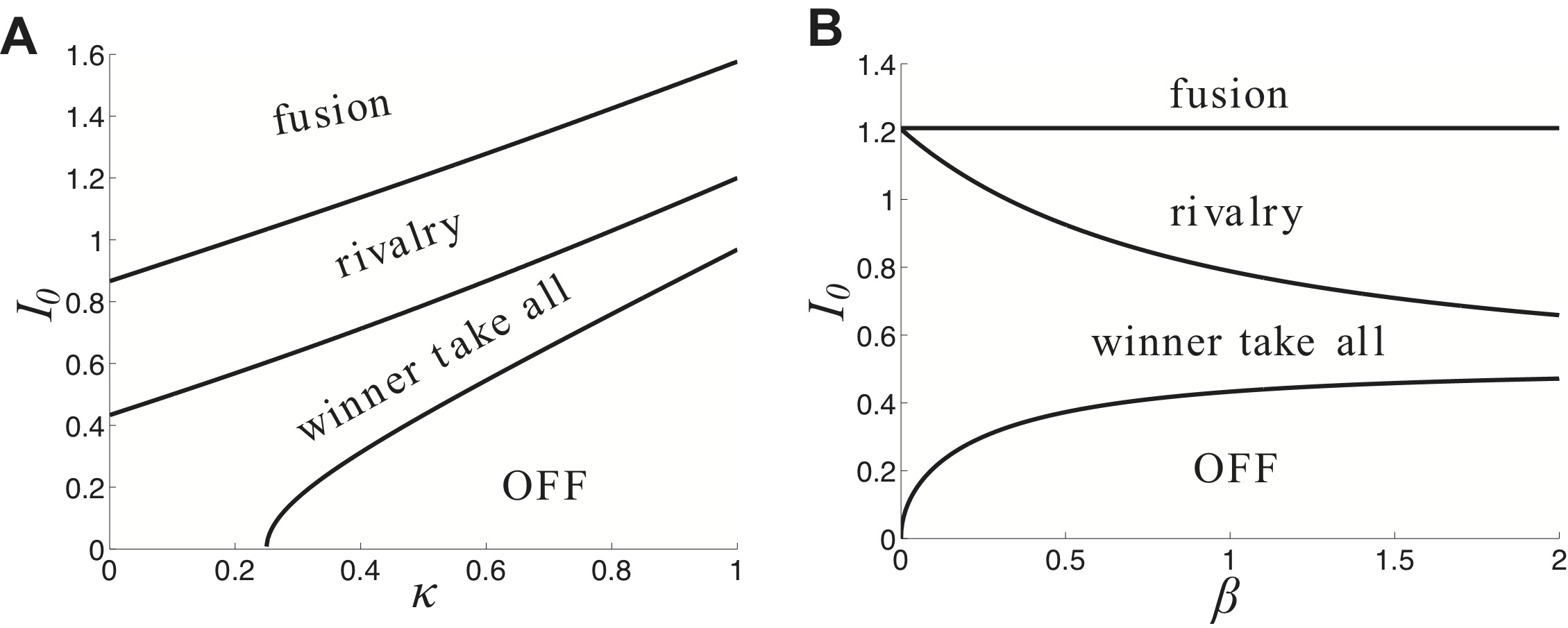} \end{center}
\caption{Partitions of parameter space into various stimulus induced states of (\ref{ringdet}). {\bf (A)} Plotted as a function of network threshold $\kappa$ and strength $I_0$ of the bimodal input (\ref{bimod}) when $\beta = 1$. {\bf (B)} Plotted as a function of synaptic depression strength $\beta$ and strength $I_0$ of bimodal input (\ref{bimod}) when $\kappa = 0.5$. Parameter $\tau = 50$.}
\label{depparsp}
\end{figure}

{\bf Fusion state.} It has been observed in many experiments on ambiguous stimuli that sufficiently strong contrast rivalrous stimuli can be perceived as a single fused image \cite{blake89,buckthought08}. This should not be surprising, considering stereoscopic vision and audition behave in exactly this way \cite{wolfe86}. However, the contrast necessary to evoke this state with dissimilar images is much higher than with similar images \cite{blake02}. In the network (\ref{ringdet}), the fusion state (Fig. \ref{rdstates}{\bf (C)}) is represented as two disjoint bumps. Therefore
\begin{align*}
U(x) = \frac{1}{1+ \beta} \left[ \int_{-\pi/ 4 -b}^{- \pi / 4 + b} + \int_{\pi / 4 - a}^{\pi / 4 + a} \right] \cos (2(x-y)) \d y - I_0 \cos (4x) + I_a \sin (2x). 
\end{align*}
Computing the integral terms, we find
\begin{align}
U(x) = \frac{{\mc S}(x,a) - {\mc S}(x,b)}{1+\beta} - I_0 \cos (4x) + I_a \sin (2x),  \label{fusasol}
\end{align}
where ${\mc S}(x,y) = \sin^2(x+y) - \sin^2(x-y)$. The solution can be specified by requiring the threshold conditions $U( - \pi / 4 \pm b) = U( \pi / 4 \pm a) = \kappa$ are satisfied
\begin{align}
\frac{\cos (2a) [ \sin (2a) - \sin (2b)]}{1+ \beta} + I_0 \cos (4a) + I_a \cos (2a) = \kappa, \label{fusaw} \\
\frac{\cos (2b) [ \sin (2b) - \sin (2a)]}{1+ \beta} + I_0 \cos (4b) - I_a \cos (2b) = \kappa, \label{fusab}
\end{align}
which we can solve numerically to relate the asymmetry of inputs $I_a$ to the half-widths $a,b$ of each bumps. In the the case of symmetric inputs, $U(x) = -I_0 \cos (4x)$, and it is straightforward to find the two bump widths explicitly. We will now study rivalrous oscillations by simply constructing them using a fast-slow analysis. We can also numerically identify the boundary of various behaviors of (\ref{ringmod}), as shown in Fig. \ref{depparsp}.

{\bf Rivalrous oscillations.} Oscillations can occur, where the two bump locations trade dominance successively (Fig. \ref{rdstates}{\bf (B)}). As in Levelt's proposition (i), increasing the contrast of a stimulus leads to that stimulus being in dominance longer. This information is not revealed when the system is stuck in a winner-take-all state. Therefore, the introduction of synaptic depression into the system (\ref{ringdet}) leads to an increase in information transfer. We will also examine how well (\ref{ringdet}) recapitulates Levelt's other propositions concerning the mean dominance of percepts.

\begin{figure}
\begin{center} \includegraphics[width=13cm]{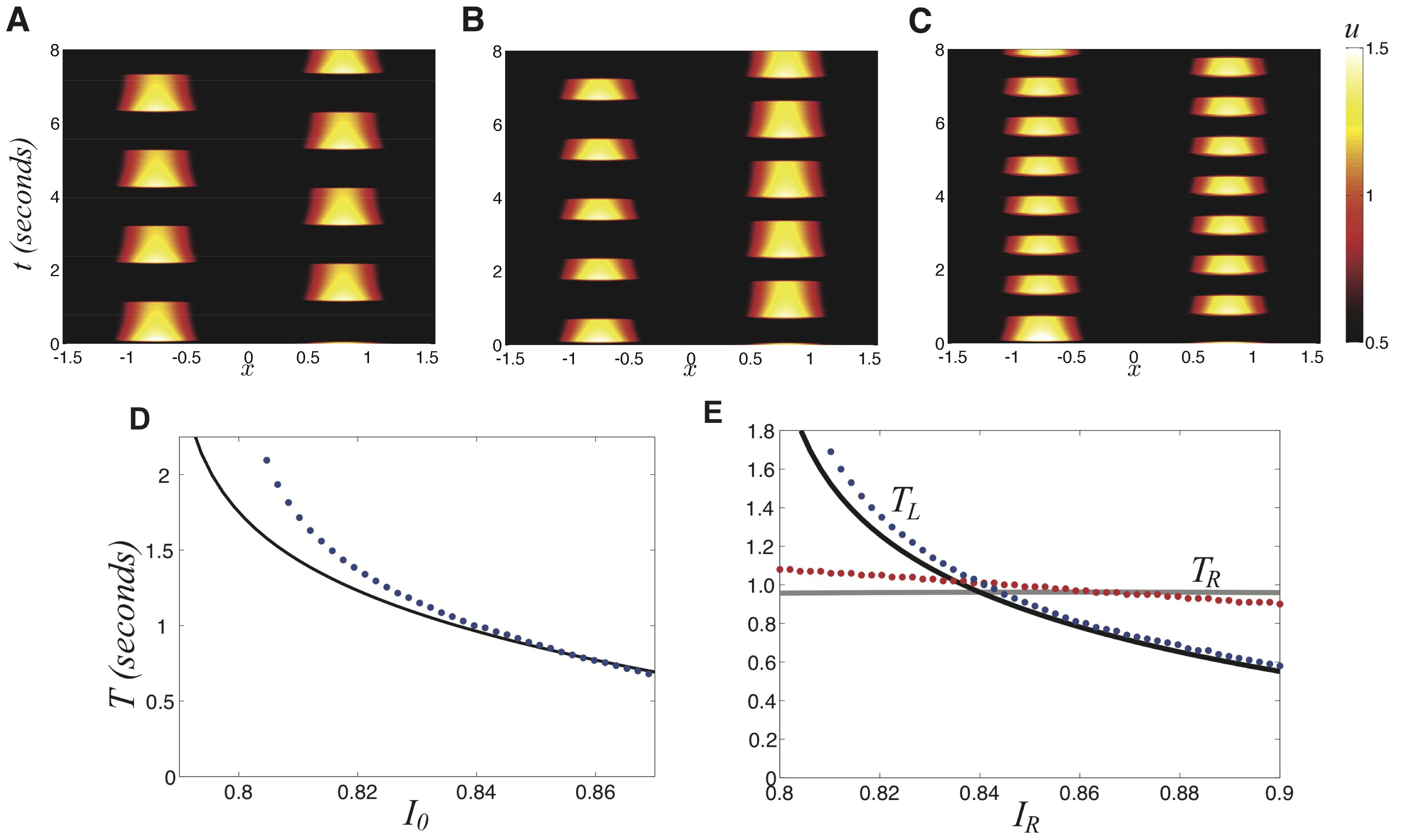} \end{center}
\caption{Dependence of rivalry on the amplitudes of the bimodal input (\ref{bimod}). {\bf (A)} Dominance times are both $T \approx 1$s when input is symmetric ($I_0 = 0.84$, $I_a =0$). {\bf (B)} Dominance time of right input ($T_R \approx 0.9$s) is longer than left ($T_L \approx 0.6$s) for asymmetric input ($I_R = 0.9$, $I_L = 0.84$). Notice $T_R$ is unchanged from case {\bf (A)}. {\bf (C)} Dominance times are both shorter of higher contrast symmetric stimulus ($I_0 = 0.9$, $I_a = 0$). {\bf (D)} Increasing the strength of the symmetric ($I_a = 0$) bimodal input (\ref{bimod}) decreases the dominance time $T$ of both populations. Our theory (black) computed from fast-slow analysis (\ref{rivdt}) fits results of numerical simulations (blue) well. {\bf (E)} For asymmetric inputs ($I_a \neq  0$), we find that varying $I_R = I_0 + I_a$ while keeping $I_L = I_0 - I_a$ fixed changes the dominance times of the left percept $T_L$ (blue) much more than that of the right percept $T_R$ (red). Other parameters are $\kappa= 0.5$, $\beta = 1$, and $\tau = 50$.}
\label{rdriv}
\end{figure}

To analyze (\ref{ringdet}) for oscillations, we assume that the timescale of synaptic depression $\tau \gg 1$, long enough that we can decompose (\ref{ringdet}) into a fast and slow system \cite{laing02,kilpatrick10b}. Synaptic input $u$ then tracks the slowly varying state of the synaptic scaling term $q$. We also assume that $q$ is essentially piecewise constant in space, in the case of the Heaviside nonlinearity (\ref{H}), which yields
\begin{align}
u (x,t) &\approx \int_{-\pi/2}^{\pi/2} \cos (2(x-y)) q(y, t) H(u(y,t)-\kappa) \d y - I_0 \cos (4x),  \label{rivoufast}
\end{align}
and $q$ is governed by (\ref{ringdet}b). To start, we will also assume a symmetric bimodal input. This way, we can simply track $q$ in the interior of one of the bumps, given $q_i(t) = q(\pi/4,t) $. Assuming a switch has just occurred, where the left bump has escaped suppression, to pin down the right, so
\begin{align}
\tau \dot{q}_i(t) = 1 - q_i(t), \ \ \ \ \ \  t \in (0,T), \ \ \ \ q_i(0) = q_0,  \label{rivqfall}
\end{align}
where $T$ is the amount of time each percept is in dominance, and $q_0$ is the synaptic strength within a bump region immediately prior to its shutting off. After the right bump escapes dominance of the left bump
\begin{align}
\tau \dot{q}_i(t) = 1 - q_i(t) - \beta q_i(t), \ \ \ \ \ \ \ t \in (T,2T).  \label{rivqrise}
\end{align}
Solving (\ref{rivqfall}) and (\ref{rivqrise}) simultaneously, we have 
\begin{align*}
q_0 = \frac{1}{1+\beta} + \frac{\beta}{1+\beta} \e^{-T/\tau} - (1-q_0) \e^{-2T/\tau},
\end{align*}
which can be solved explicitly for the dominance time
\begin{align}
T &= \tau \ln \left[ \frac{\beta + \sqrt{\beta^2 - 4 (1 + \beta ) (1-q_0) [(1+\beta) q_0 -1]} }{2(1+\beta)q_0 - 2} \right],  \label{rivdt}
\end{align}
so that we now must specify the value $q_0$. We can examine the fast equation (\ref{rivoufast}), solving for the form of the slowly narrowing right bump during its dominance phase
\begin{align}
u(x,t) &= q_i(t) \int_{\pi/4-a(t)}^{\pi/4+a(t)} \cos (2(x-y)) \d y - I_0 \cos (4x) \nonumber \\
&= q_i(t) \left[ \sin^2(x+a(t)) - \sin^2 (x - a(t)) \right] - I_0 \cos (4x).  \label{uslowbump}
\end{align}
We can solve for the slowly evolving width $a(t)$ of this bump by requiring the threshold condition $u(\pi/4 \pm a(t), t) = \kappa$ to yield
\begin{align*}
\frac{q_i(t)}{2} \sin (4 a(t)) + I_0 \cos (4 a(t)) = \kappa,
\end{align*}
and then solving using trigonometric identities
\begin{align}
a(t) = \frac{1}{2} \tan^{-1} \left[ \frac{q_i(t) + \sqrt{q_i(t)^2 + 4 (I_0^2 - \kappa^2)}}{2(I_0 + \kappa)} \right].  \label{rivaslow}
\end{align}
We can also identify the maximal value of $q_i(t) = q_0$ which still leads to the right bump suppressing the left. Once $q_i(t)$ falls below $q_0$, the other bump escapes suppression, flipping the dominance of the current bump. This is the point at which the other bump of (\ref{uslowbump}) rises above threshold, as defined by the equation
\begin{align*}
u( - \pi/4, t) = I_0 - q_0 \sin (2 a_0) = \kappa.
\end{align*}
Combining this with the equation (\ref{rivaslow}), we have an algebraic equation for $q_0$ given
\begin{align*}
(I_0-\kappa)^2 = q_0^2 \frac{2q_0^2+4(I_0^2-\kappa^2) + 2 q_0 \sqrt{q_0^2 + 4(I_0^2-\kappa^2)}}{8I_0^2 + 8I_0\kappa + 2 q_0^2 + 2 q_0 \sqrt{q_0^2 + 4(I_0^2- \kappa^2)}},
\end{align*}
which is straightforward to solve for
\begin{align}
q_0 = \frac{2I_0 \sqrt{(I_0 - \kappa) (3 I_0 + \kappa)}}{3I_0 + \kappa}  \label{q0}
\end{align}
and we have excluded an extraneous negative solution. Interestingly, the amplitude of synaptic depression is excluded from (\ref{q0}), but we do know based on (\ref{rivqfall}) and (\ref{rivqrise}) that $q_0 \in ([1+\beta]^{-1},1)$. This establishes a bounded region of parameter space in which we can expect to find rivalrous oscillations, which we use to construct a partitioning of parameter space in Fig. \ref{depparsp}. We can also now approximate the dominance time using (\ref{rivdt}) with (\ref{q0}), as shown in Fig. \ref{rdriv}{\bf (D)}.

In the case of an asymmetric bimodal input ($I_a>0$), we can also solve for explicit approximations to the dominance times of the right $T_R$ and left $T_L$ populations. Following the same formalism as for the symmetric input case
\begin{align}
T_R & = \tau \ln \left[ \frac{\beta + q_d + \sqrt{(\beta + q_d)^2 - 4(1+ \beta) (1 - q_L)[(1+ \beta)q_R - 1]}}{2 (1+ \beta) q_R - 2} \right], \label{rivTR} \\
T_L &= \tau \ln \left[ \frac{\beta - q_d + \sqrt{(\beta - q_d)^2 - 4(1+ \beta) (1- q_R)[(1+ \beta) q_L - 1]}}{2 (1+ \beta) q_L - 2} \right], \label{rivTL}
\end{align}
where $q_d = (1+ \beta ) (q_R - q_L)$, in terms of the local values $q_L$ and $q_R$ of the synaptic scaling in the right and left bump immediately prior to their suppression. Notice in the case $q_L = q_R$, then $q_d=0$ and (\ref{rivTR}) and (\ref{rivTL}) both reduce to (\ref{rivdt}). We now need to examine the fast equation (\ref{rivoufast}) to identify these two values. This is done by generating two implicit equations for the half-width $a_R$ and $q_R$ at the time of a switch
\begin{align}
\frac{q_R}{2} \sin (4 a_R) + I_0 \cos (4 a_R) + I_a \cos (2a_R) & = \kappa, \\
I_0 - I_a - q_R \sin (2a_R) & = \kappa, 
\end{align}
which we can solve explicitly for
\begin{align}
a_R &= \frac{1}{2} \cos^{-1} \left[ \frac{\kappa}{2 I_0} + \frac{1}{2} \right],
\end{align}
and
\begin{align}
q_R &= \frac{2 I_0 (I_L - \kappa)}{\sqrt{(3I_0 + \kappa)(I_0 - \kappa)}}, \label{rivqR}
\end{align}
where $I_L = I_0 - I_a$ is the strength of input to the left side of the network. Likewise, we can find the value of the synaptic scaling in the left bump immediately prior to its suppression
\begin{align}
q_L &= \frac{2 I_0 (I_R - \kappa)}{\sqrt{(3 I_0 + \kappa)(I_0 - \kappa)}}, \label{rivqL}
\end{align}
where $I_R = I_0 + I_a$ is the strength of input to the right side of the network. Using the expressions (\ref{rivqR}) and (\ref{rivqL}) we can now compute the dominance time formulae (\ref{rivTR}) and (\ref{rivTL}), showing the relationship between inputs and dominance times in Fig. \ref{rdriv} {\bf (E)}. Notice that all of Levelt's propositions are essentially satisfied. Changing the strength of the right stimulus $I_R$ has a very weak effect on the dominance time of the right percept. However, it does increase the overall alternation rate and decrease the proportion of time the left percept remains in dominance.

We also note there is a critical strength of synaptic depression $\beta$ below which rivalrous oscillations do not occur. When synaptic depression is sufficiently strong, the winner-take-all state ceases to exist. Beyond this critical synaptic depression strength, the network either supports rivalrous oscillations or a fusion state. Either way, information is conveyed to the network that would otherwise be kept hidden. We show this in Fig. \ref{depparsp}{\bf (B)}. In this way, synaptic depression can improve the information transfer of the network (\ref{ringdet}). In fact, we will show that it does so in a way that is much more reliable than noise.

\subsection{Purely stochastic switching in the ring model}

We will now study rivalrous switching brought about by fluctuations. In particular, we ignore depression and examine the noisy system
\begin{align}
\d u(x,t) = \left[ - u(x,t) + \int_{- \pi / 2}^{\pi / 2} w (x-y) f(u(y,t)) \d y + I (x) \right] \d t  + \d \xi (x,t).  \label{unoise}
\end{align}
where $\langle \xi (x,t) \rangle = 0$ and $\langle \xi (x,t) \xi (y,s) \rangle = \ve C(x-y) \delta (t-s)$ defines the spatiotemporal correlations of the system.

\begin{figure}
\begin{center} \includegraphics[width=13cm]{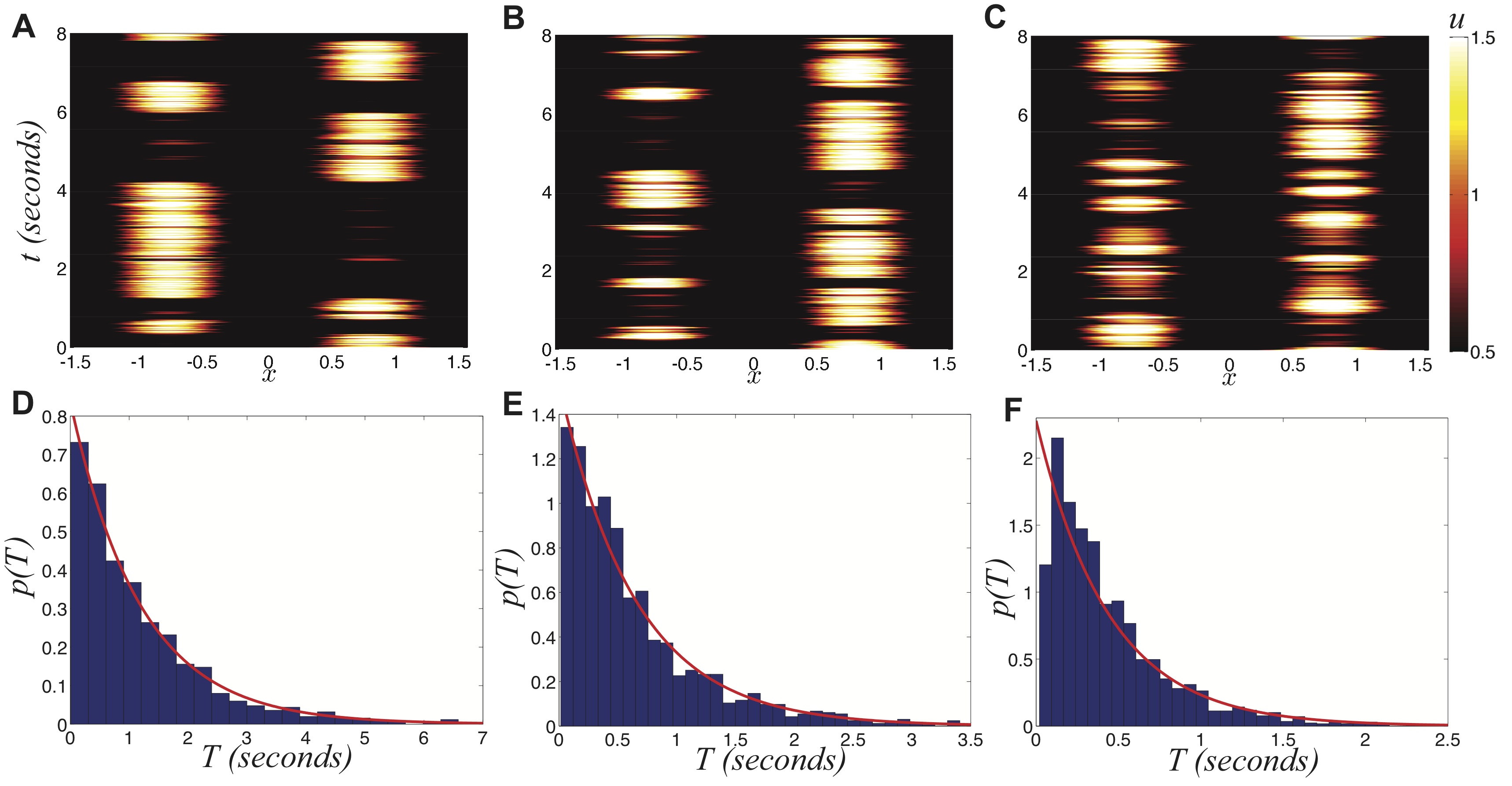}  \end{center}
\caption{Purely noise induced switching of dominance in the depression-free ring model (\ref{unoise}) {\bf (A)} Numerical simulations of the system for the various input strengths of the symmetric ($I_a = 0$) bimodal input (\ref{bimod}) with {\bf (A)} $I_0 = 0.8$, {\bf (B)} $I_0 = 0.9$, and {\bf (C)} $I_0 = 1.0$. Distributions of dominance times computed numerically (blue bars) with the exponential distribution (\ref{switchexp}) with numerically computed mean $\langle T \rangle$ (red) superimposed for {\bf (D)} $I_0 = 0.8$ has $\langle T \rangle \approx 1.2$s, (e) $I_0 =0.9$ has $\langle T \rangle \approx 0.70$s, and (f) $I_0 = 1.0$ has $\langle T\rangle \approx 0.45$s. Other parameters are $\kappa = 0.5$ and $\ve = 0.04$.}
\label{uswitch}
\end{figure}

To start, we can simply consider (\ref{unoise}) in the absence of noise
\begin{align}
u_t (x,t) = - u(x,t) + \int_{- \pi / 2}^{\pi / 2} w (x-y) f(u(y,t)) \d y + I(x).  \label{nodepmod}
\end{align}
This model has been studied extensively \cite{amari77,benyishai95}, so we will not perform an in depth analysis of stationary bump solutions. We are interested in the winner-take-all state. For a cosine weight (\ref{cos}), Heaviside firing rate (\ref{H}), and bimodal stimulus (\ref{bimod}) these can be computed as a special case of (\ref{wtasol}) and lie at either $x= \pm \pi / 4$, so
\begin{align}
U_{\pm}(x) = \sin (2a_{\pm}) \sin (2x) - I_0 \cos (4x) \pm I_a \sin (2x),  \label{wtaring}
\end{align}
and we can apply the threshold condition $U_{\pm}(  \pm \pi/4 + a_{\pm} ) = \kappa$, so
\begin{align}
\frac{1}{2} \sin (4a_{\pm}) + I_0 \cos (4a_{\pm}) \pm I_a \cos (2a_{\pm}) = \kappa. \label{wtand}
\end{align}
In the case of a symmetric input, $a_{\pm}  = a$ and we can solve (\ref{wtand}) explicitly
\begin{align}
a = \frac{1}{2} \tan^{-1} \left[ \frac{1 + \sqrt{1 + 4(I_0^2 - \kappa^2)}}{2(I_0 + \kappa)} \right].
\end{align}
Since we have no synaptic depression in the model (\ref{nodepmod}), we cannot rely on deterministic mechanisms to generate switches between one winner-take-all state and another. Therefore, we will consider the effects of introducing a small amount of noise ($0 < \ve \ll 1$), reflective of synaptic fluctuations. We focus on the spatial correlation function $C(x) = \cos (x)$. Noise can generate switches in between the two dominant states (Fig. \ref{uswitch}). As the strength of both input contrasts are increased, switches between percepts occur more often. In fact, there is an exponential dependence of the mean dominance time $\langle T \rangle$ on the strength $I_0$ of the bimodal input (\ref{bimod}). We will provide an argument, using energy methods, as to why this occurs in our analysis of the simplified system (\ref{fulcompnet}).

\begin{figure}
\begin{center} \includegraphics[width=13cm]{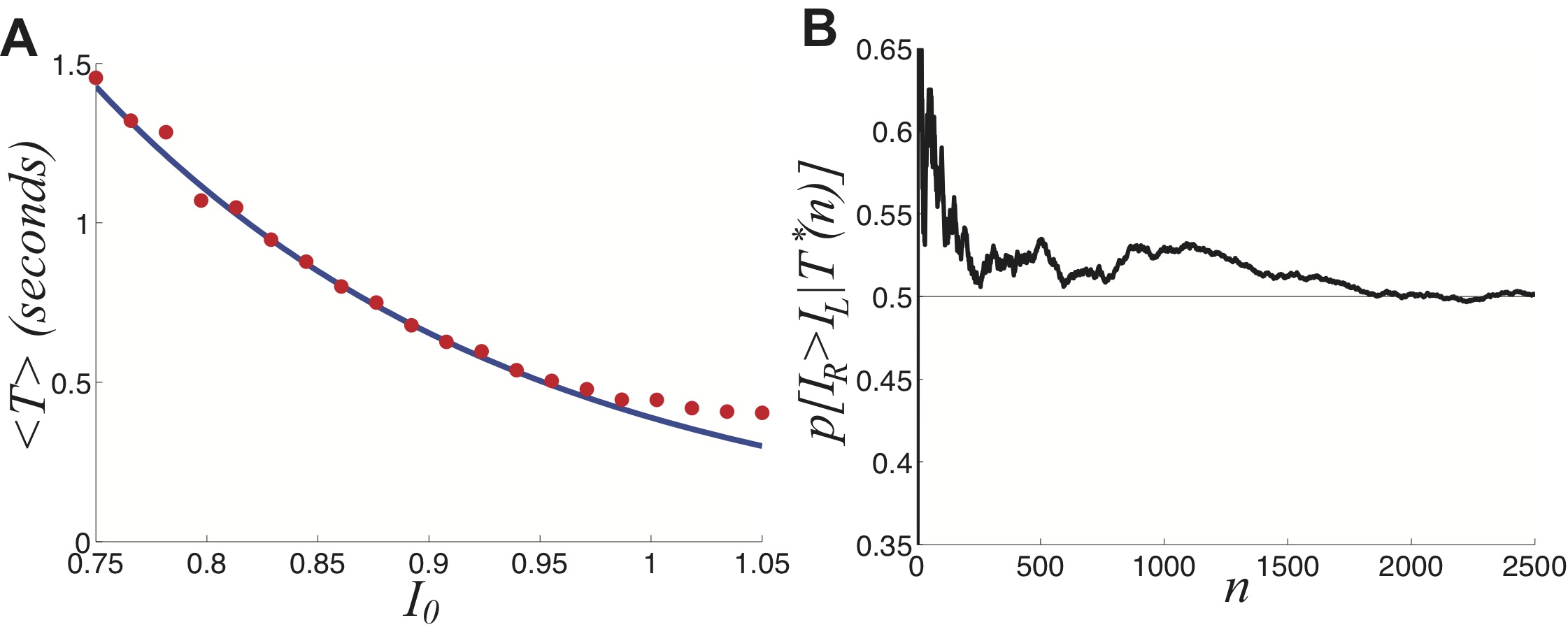} \end{center}
\caption{Sampling of the strength of each input based on noise-induced transitions. {\bf (A)} Mean dominance time $\langle T \rangle$ as a function of the strength $I_0$ of the symmetric ($I_a = 0$) bimodal input (\ref{bimod}). {\bf (B)} Probability that the right input $I_R$ was higher than the left output $I_L$, based on the sampling $n$ cycles ($2n$ switches between percepts), in the case of symmetric inputs $I_L = I_R = 0.9$. Notice it takes close to 2000 cycles before $p[I_R>I_L | T^*(n)] \approx 0.5$. Other parameters are $\kappa = 0.5$ and $\ve = 0.04$.}
\label{usym}
\end{figure}

We will study switching in the case that the suppressed bump comes on and shuts off the bump that is currently on. This mechanism is known as escape \cite{wang92}. As shown in Fig. \ref{uswitch}{\bf (A-C)}, dominance times decrease with input strength on average. Here escape is a noise induced effect \cite{moreno07}, rather than a deterministic depression-induced effect \cite{shpiro07,kilpatrick10b}. However, as opposed to depression-induced switching, there is a substantial spread in the possible dominance times for a given set of parameters (Fig. \ref{uswitch}{\bf (D-F)}). Thus, by examining two dominance times back to back, an observer would have difficulty telling if the input strengths were roughly the same or not. Recently, psychophysical experiments have been carried out where an observer must identify the higher contrast of two rivalrous percepts \cite{moreno11}, showing humans perform quite well in at this task. Results of \cite{moreno11} suggest humans most likely use Bayesian inference in discerning information about visual percepts. This is in keeping with previous observations that humans' visual perception of objects likely carries out Bayesian inference \cite{kersten04}.

Notice in Fig. \ref{usym}{\bf (B)} that the likelihood an observer assigns to $I_R>I_L$ approaches 1/2 as the number of observations $n$ increases. We compute $p[I_R>I_L | T^*(n)]$, the probability an observer would presume $I_R>I_R$ conditioned on dominance time pairs from $n$ cycles $T^*(n) = \left\{ T^{(1)}_R, T^{(1)}_L, T^{(2)}_R, T^{(2)}_L, ... , T^{(n)}_R, T^{(n)}_L  \right\} $, numerically here. However, as the number of cycles $n \to \infty$, the exponential distributions approximately defining the identical probability densities $p_R(T_R) = p_L(T_L) = p(T)$ will be fully sampled. We can calculate
\begin{align*}
p(I_R > I_L | T^* ( \infty ) ) = \int_0^{\infty} \int_0^x p(x) p(y) \d y \d x = \frac{1}{\langle T \rangle^2} \int_0^{\infty} \int_0^x \e^{-(x+y)/ \langle T \rangle } \d y \d x = \frac{1}{2},
\end{align*}
as in Fig. \ref{usym}{\bf (B)}. In the case where depression, in the absence of noise, drives switches, we would expect this limit to be approached much more quickly. We will also examine how a combination of depression and noise affects an observer's ability to discern contrast differences.

\begin{figure}
\begin{center} \includegraphics[width=12cm]{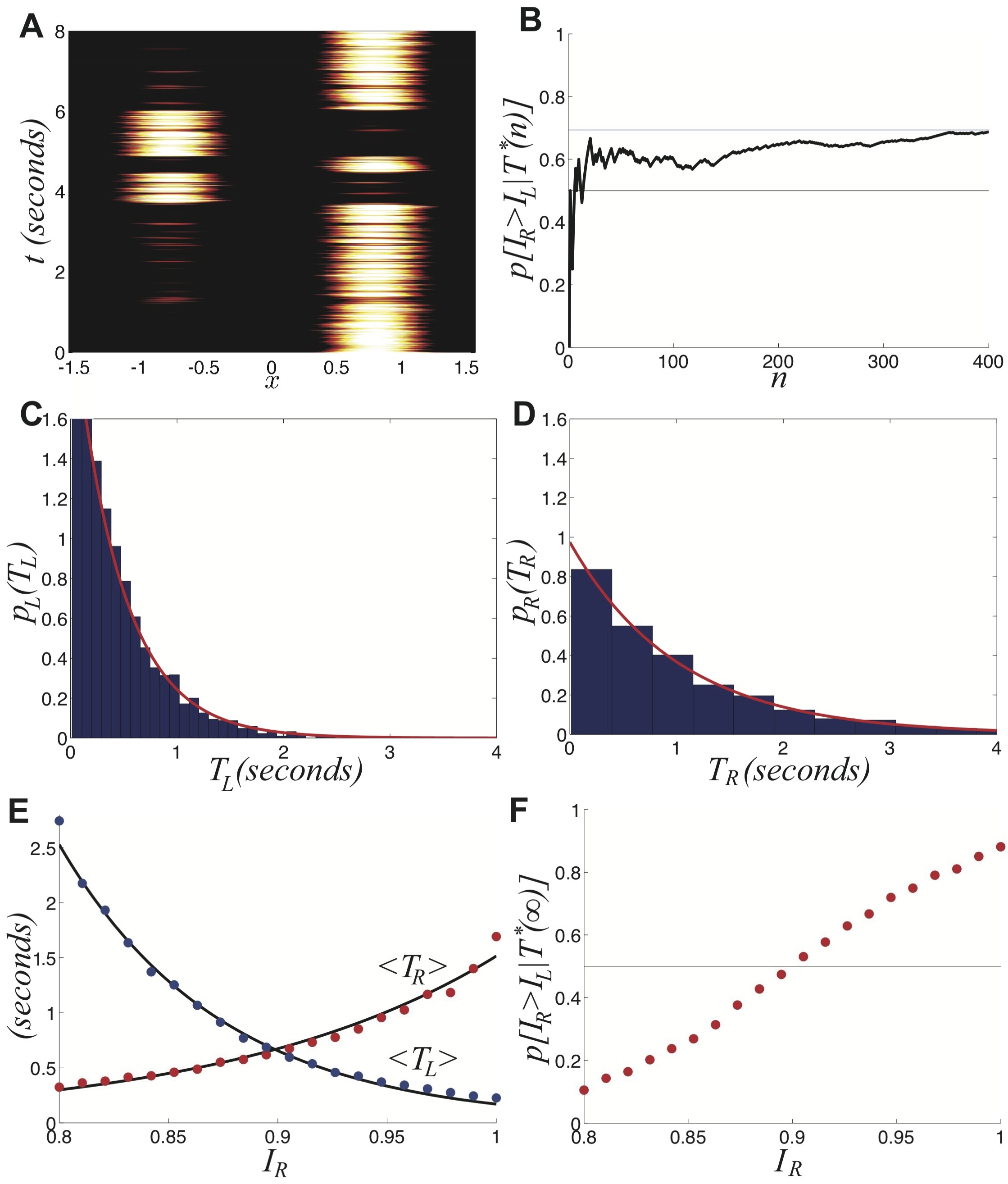}    \end{center}
\caption{{\bf (A)} Single realization of the stochastic neural field (\ref{unoise}) with asymmetric ($I_a >0$) inputs $I_R = 0.92$ and $I_L = 0.88$, leads to longer dominance times for right percept $T_R$. {\bf (B)} Expected likelihood $p[I_R>I_R|T^*(n) ]$ that the right input $I_R$ is stronger than left $I_L$ based on $n$ comparisons of dominance times $T_R$ and $T_L$ sampled. Blue line is theoretical prediction (\ref{Tmeanrat}) of the limit as $n \to \infty$. Numerically computed dominance time distributions (blue bars) are well fit by the exponential distribution (\ref{switchexp}) for the {\bf (C)} left ($\langle T_L \rangle \approx 0.5$s) and {\bf (D)} right ($\langle T_R \rangle \approx 1$s) percepts. {\bf (E)}Dependence of mean dominance times $\langle T_R \rangle$ and $\langle T_L \rangle$ on the strength of the right input $I_R$. Black curves are best fits to exponential functions of $I_R$. {\bf (F)} Expected likelihood $p[I_R>I_L | T^* ( \infty )]$ right input $I_R$ is stronger than left $I_L$ in the limit of high sample number $n \to \infty$, as computed theoretically by (\ref{Tmeanrat}). Other parameters are $\kappa = 0.5$, and $\ve = 0.04$.}
\label{uswasym}
\end{figure}

We explore this further in the case of asymmetric inputs, showing that dominance times are still specified by roughly exponential distributions as shown in Fig. \ref{uswasym}. When $I_R> I_L$, even though the means satisfy $\langle T_R \rangle > \langle T_L \rangle$ (Fig. \ref{uswasym}{\bf (E)}), the exponential distributions $p(T_R)$ and $p(T_L)$ have considerable variance, also given by the means. Therefore, randomly sampled values from these distributions may satisfy $T_R<T_L$. Were an observer to use one such sample as a means for guessing the inputs that generated them, they would guess $I_R<I_L$, rather than the correct $I_R>I_L$. In terms of conditional probabilities, we can expect situations where $p(I_R>I_L | T^*(n) ) < 1/2$ for finite $n$, even though $I_R>I_L$. We can quantify this effect numerically, as shown in Fig. \ref{uswasym}{\bf (F)}. In the limit $n \to \infty$, we find uncertainty continues to creep in, since fluctuations continually give an observer misleading information. Since the marginal distributions are approximately exponential
\begin{align}
p_j (T_j) =  \e^{- T_j / \langle T_j \rangle }/ \langle T_j \rangle \ \ \ \ j=L,R,  \label{switchexp}
\end{align}
we can approximate the conditional probability
\begin{align}
p[I_R>I_L | T^*(\infty)] &= \int_0^{\infty} \int_0^x p_R(x) p_L(y) \d y \d x \nonumber \\
&= \frac{1}{\langle T_R \rangle \rangle T_L \rangle} \int_0^{\infty} \int_0^x \e^{- x/ \rangle T_R \rangle} \e^{- y/ \rangle T_L \rangle} \d y \d x \nonumber \\
&= 1 - \frac{\langle T_L \rangle}{\langle T_R \rangle + \langle T_L \rangle} = \frac{\langle T_R \rangle}{\langle T_R \rangle + \langle T_L \rangle}. \label{Tmeanrat}
\end{align}
Observe that the approximation we make using the formula (\ref{Tmeanrat}) accurately estimates the limit $p(I_R>I_L| T^* (\infty))$ as shown in Fig. \ref{uswasym}{\bf (B)}. This is the likelihood that an observer performing Bayesian sampling of the probability densities (\ref{switchexp}) will predict $I_R > I_L$. Recent psychophysical experiments suggests humans would perform this task of contrast differentiation in this way \cite{kersten04,moreno11}.

We see from our analysis that when switches are generated by noise, rather than deterministic depression, the means dominance times still obey Levelt's propositions to some extent (Fig. \ref{uswasym}{\bf (E)}). This would allow for an accurate comparison of input strengths $I_R$ and $I_L$ based on the means $\langle T_R \rangle$ and $\langle T_L \rangle$. However, when considering a more realistic observer, that could only compare successive dominance times, accurately discerning the comparison of the input contrasts is more difficult. This becomes much more noticeable when the input contrasts are quite close to one another, as in Fig. \ref{uswasym}{\bf (F)}. We will explore now how introducing depression along with noise improves discernment of the input contrasts by an observer using simple comparison of dominance times.

\subsection{Switching through combined depression and noise}

\begin{figure}
\begin{center}  \includegraphics[width=13cm]{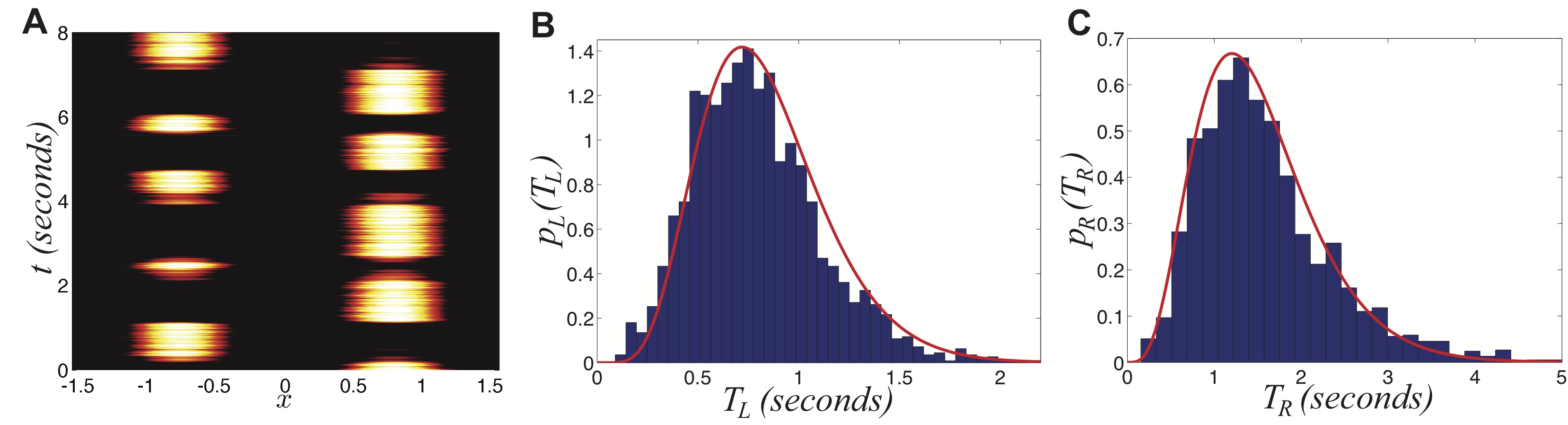} \end{center}
\caption{Switching in the stochastic ring model with depression (\ref{ringmod}) with asymmetric inputs ($I_a>0$). {\bf (A)} Single realization for asymmetric inputs with $I_R = 0.92$ and $I_L = 0.88$, which leads to right percept dominating longer. {\bf (B)} Distribution of left percept dominances times $p_L(T_L)$ over 1000s is well fit by a gamma distribution (\ref{gamdist}). {\bf (C)} Distribution of right percept dominance times $p_R(T_R)$ across 1000s is well fit by a gamma distribution (\ref{gamdist}). Other parameters are $\kappa = 0.5$, $\beta = 0.2$, $\tau = 50$, and $\ve = 0.01$.}
\label{uswqon}
\end{figure}

\begin{figure}
\begin{center} \includegraphics[width=13cm]{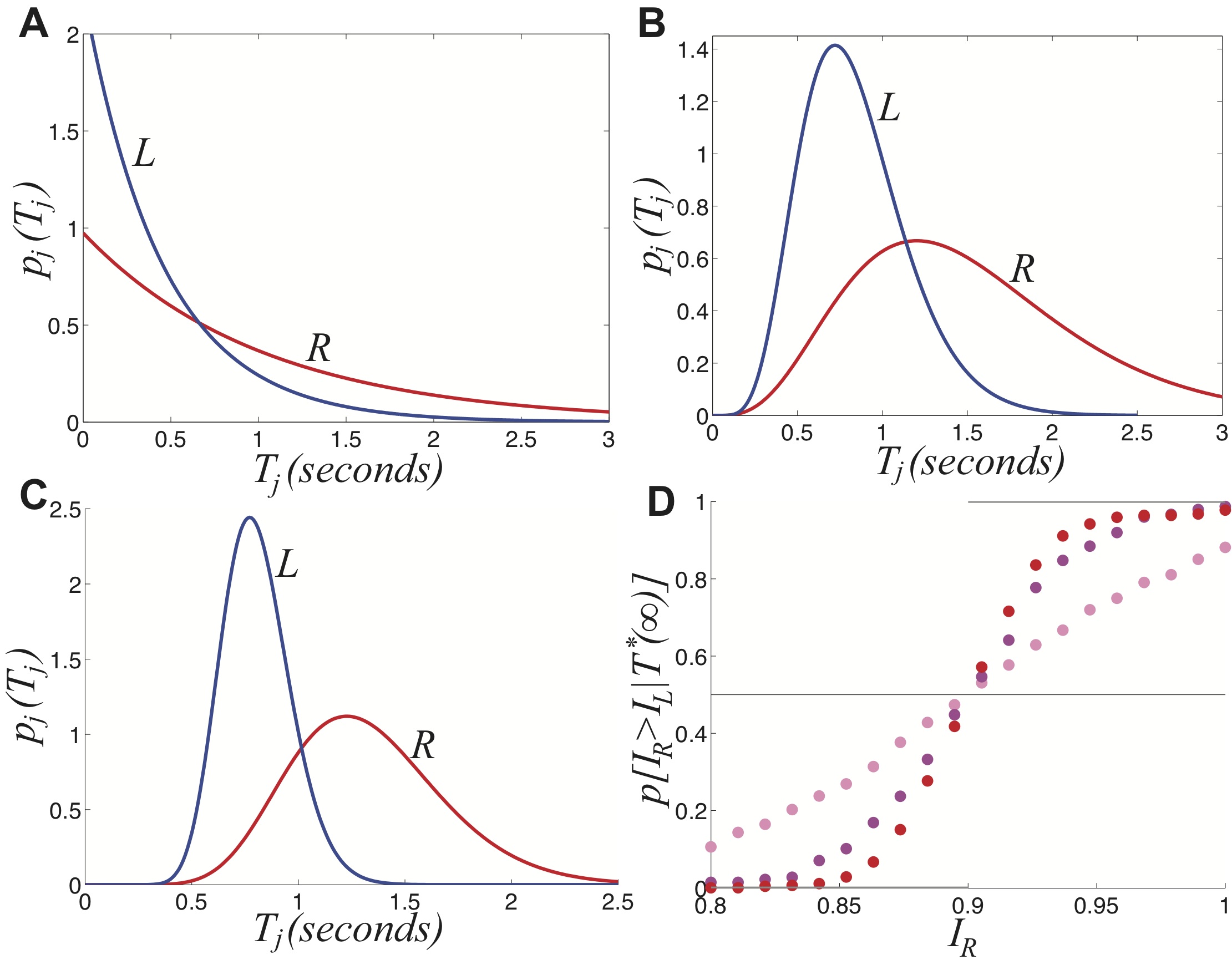} \end{center}
\caption{Comparing the probability densities of dominance times in the stochastic ring model with depression (\ref{ringmod}) for various levels of noise and depression. {\bf (A)} No depression and $\ve = 0.04$ {\bf (B)} Depression strength $\beta = 0.2$ and $\ve = 0.01$. {\bf (C)} Depression strength $\beta = 0.4$ and $\ve = 0.0025$. {\bf (D)} Expected likelihood $p[I_R>I_R|T^*( \infty)]$ the right input $I_R$ is stronger than the left $I_L$ based in the limit of an infinite number of samples of the dominance times $T_R$ and $T_R$ for the parameters in {\bf (A)} (pink); {\bf (B)} (magenta); and {\bf (C)} (red). Other parameters are $\tau = 50$ and $\kappa = 0.5$.}
\label{pcompring}
\end{figure}

We now study the effects of combining noise and depression in the full ring model of perceptual rivalry (\ref{ringmod}). Numerical simulations of (\ref{ringmod}) reveal that noise-induced switches occur robustly, even in parameter regimes where the noise-free system supports no rivalrous oscillations, as shown in Fig. \ref{uswqon}. Rather than dominance times being distributed exponentially, they roughly follow a gamma distribution \cite{fox67,lehky95}
\begin{align}
p_j (T_j)  = \frac{1}{\sigma^k \Gamma (k)} T_j^k \exp \left[ - T_j/ \sigma \right], \ \ \ \  k > 1.  \label{gamdist}
\end{align}
As opposed to the exponential distribution, (\ref{gamdist}) is peaked away from zero at $T_j = k \sigma$, which is also the mean of the distribution.  Therefore, two distributions of dominance times with different means will be more easily discerned from one another. We show this in Fig. \ref{pcompring}{\bf (B)} by superimposing the two distributions from Fig. \ref{uswqon} on top of one another. They clearly separate better than in the probability densities of purely noise-driven switching, shown in Fig. \ref{pcompring}{\bf (A)}. As the strength of synaptic depression is increased even further, keeping the mean dominance times $\langle T_R \rangle$ and $\langle T_L \rangle$ the same, probability densities separate even further (Fig. \ref{pcompring}{\bf (C)}). We summarize how this separation improves the inference of input contrast difference in Fig. \ref{pcompring}{\bf (D)}. As the strength $\beta$ of depression is increased and noise is decreased, an observer's ability to discern which input was stronger is improved. The likelihood assigned to $I_R$ being greater than $I_L$ is a sigmoidal function of $I_R$ whose steepness increases with $\beta$. For no noise, the likelihood function is simply a step function $H(I_R>I_L)$, implying perfect discernment.

\subsection{Analyzing switching in a reduced model}

\begin{figure}
\begin{center}  \includegraphics[width=13cm]{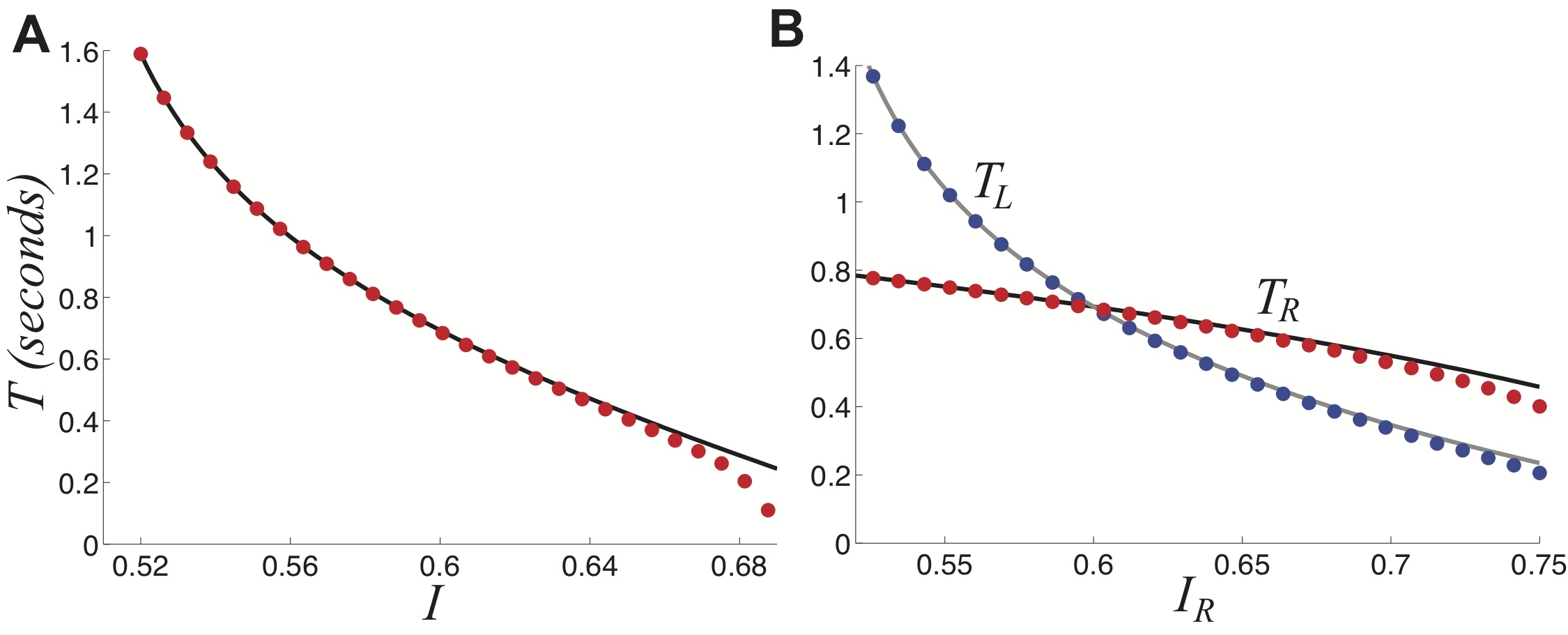} \end{center}
\caption{Dominance times computed adiabatically in the noise-free competitive network with depression (\ref{scdet}). {\bf (A)} Plot of dominance times $T$ as a function of the strength of a symmetric input $I_R = I_L = I$ to the competitive system (\ref{scdet}) show fast-slow theory (curve) match numerical simulations (dots) very well. {\bf (B)} Dominance times $T_L$ and $T_R$ as a function of right input $I_R$ keeping $I_L = 0.6$ fixed as computed by theory (curves) in (\ref{scdtTR}) and (\ref{scdtTL}) fits numerically computed (dots) very well. Other parameters are $\beta = 1$ and $\tau = 50$.}
\label{scdetf}
\end{figure}

We now perform similar analysis on a reduce competitive network model (\ref{fulcompnet}) and extend some of the results for the ring model. One of the advantages is that we can construct an energy function \cite{hopfield84}, which provides us with intuition as to the exponential dependence of mean dominance times on input strengths in the noise-driven case. In particular, we will analyze (\ref{fulcompnet}) where the firing rate function is Heaviside (\ref{H}), starting with the case of no noise
\begin{subequations}  \label{scdet}
\begin{align}
\dot{u}_R & = - u_R + H(I_R - q_L u_L), \  \ \ \ \  \dot{u}_L = - u_L + H(I_L - q_R u_R) \\
\tau \dot{q}_R &= 1 - q_R - \beta u_R q_R, \ \ \ \ \tau \dot{q}_L = 1 - q_L - \beta u_L q_L.
\end{align}
\end{subequations}
First, we note (\ref{scdet}) has a stable winner-take-all solution in the $j$th population ($j=R,L$) for $I_j>0$ and $I_k<1/(1+ \beta )$ ($k \neq j$). Second, a stable fusion state exists when both $I_L,I_R>1/(1+ \beta )$. Coexistent with the fusion state, there may be rivalrous oscillations, as we found in the spatially extended system (\ref{ringdet}). To study these, we make a similar fast-slow decomposition of the model (\ref{scdet}), assuming $\tau \gg 1$ to find $u_j$'s possess the quasi-steady state
\begin{align}
u_R = H(I_R - q_L u_L), \ \ \ \ \  \ u_L  = H(I_L - q_R u_R).  \label{uscfast}
\end{align}
so we expect $u_j = 0$ or $1$ almost everywhere. Therefore, we can estimate the dominance time of each stimulus using a piecewise equation for the slow subsystem
\begin{align}
\tau q_j = \left\{ \begin{array}{ll} 1 - q_j - \beta q_j & : u_j = 1, \\ 1-q_j & : u_j = 0, \end{array} \right. \ \  \ \ \ \ j = R,L.  \label{qscslow}
\end{align}
Combining the slow subsystem (\ref{qscslow}) with the quasi-steady state, we can use self-consistency to solve for the dominance times $T_R$ and $T_L$ of the right and left populations. We simply note that switches will occur through escape mechanism, when the cross-inhibition between populations becomes weak enough such that the suppressed population's ($j$) input becomes superthreshold, so $I_j = q_k$. Using (\ref{qscslow}) as we did in the spatial system,  we find
\begin{align}
T_R &=  \tau \ln \left[ \frac{\beta - I_d + \sqrt{(\beta - I_d)^2 - 4( 1 - I_R) ( 1+ \beta ) [(1+ \beta) I_L - 1]}}{2(1+ \beta ) I_L  - 2} \right], \label{scdtTR} \\
T_L &= \tau \ln \left[ \frac{\beta + I_d + \sqrt{(\beta + I_d)^2 - 4( 1 - I_L) ( 1+ \beta ) [(1+ \beta) I_R - 1]}}{2(1+ \beta ) I_R  - 2} \right], \label{scdtTL}
\end{align}
where $I_d = (1+ \beta ) [ I_R - I_L]$. For symmetric stimuli, $I_L = I_R = I$, both (\ref{scdtTR}) and (\ref{scdtTL}) reduce to
\begin{align}
T &= \tau \ln \left[ \frac{\beta + \sqrt{\beta^2 - 4 (1 - I) (1 + \beta )[(1+ \beta ) I - 1]}}{2(1+ \beta) I - 2} \right], \label{scdtsym}
\end{align}
using which we can solve for the critical input strength $I$ above which only the fusion state exists
\begin{align}
I = \frac{2 + \beta}{2(1+ \beta)},
\end{align}
in the case of symmetric inputs. We show in Fig. \ref{scdetf} that this asymptotic approximations (\ref{scdtTR}) and (\ref{scdtTL}) of the dominance times match well with the results of numerical simulations. Levelt's propositions are recapitulated well.

\begin{figure}
\begin{center} \includegraphics[width=13cm]{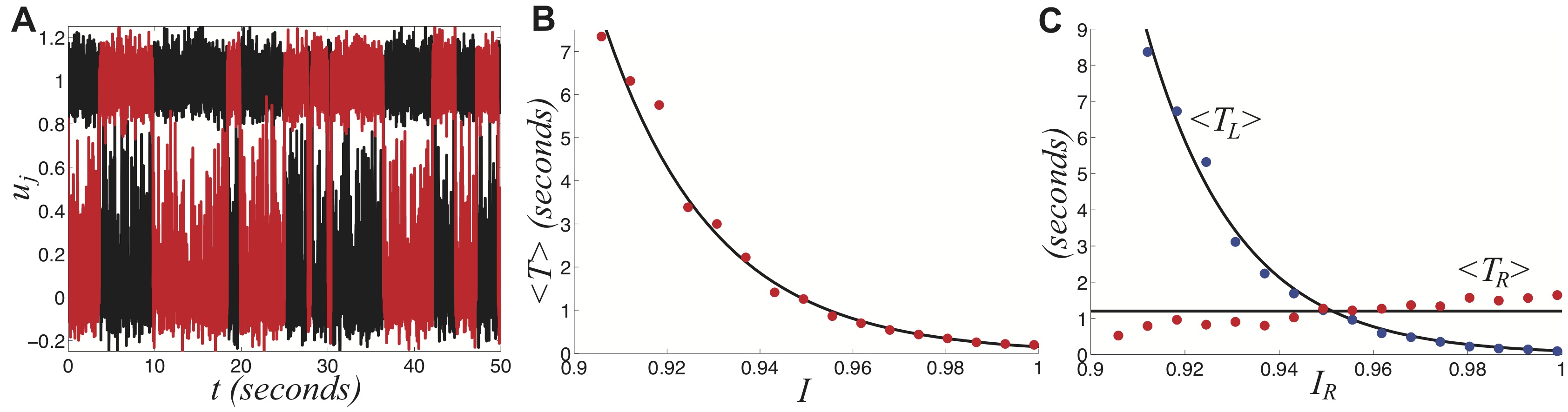} \end{center}
\caption{Noise induced transitions in depression-free two population network (\ref{compsimp}). {\bf (A)} Single realization with $I_R = I_L = I = 0.9$ for the right $u_R$ (red) and left $u_L$ (black) population activities. {\bf (B)} Mean dominance time $\langle T \rangle$ as a function of input strength $I$ computed numerically (red dots) and fit to the theoretically derived exponential function (\ref{mdtcomps}). {\bf (C)} Mean dominance times $\langle T_R \rangle$ and $\langle T_L \rangle$ as a function of the right input strength $I_R$ while $I_L = 0.95$ is fixed.  Other parameters are $\ve = 0.01$.}
\label{sccmpnois}
\end{figure}

Now, we study noise-induced switching in the competitive network. We can separate timescales to study the effects of depression and noise together. To start, we consider the limit of no depression $\beta \to 0$, so that
\begin{subequations} \label{compsimp}
\begin{align}
\dot{u}_R &= -u_R + H(I_R - u_L )  + \xi_R , \\
\dot{u}_L &= -u_L + H(I_L - u_R ) + \xi_L, 
\end{align}
\end{subequations}
where $\xi_j$ are independent white noise processes with variance $\ve$. We show a single realization of the competitive network in Fig. \ref{sccmpnois}{\bf (A)}. Most of the time, the dynamics remains close to one of the winner-take-all attractors where $u_j = 1$ and $u_k = 0$ ($j=R,L$ and $k \neq j$). Occasionally, noise causes large deviations where the suppressed population's activity rises above threshold, causing the once dominant population to then be suppressed. We plot the relationship between the strength of the inputs and the mean dominance times in Fig. \ref{sccmpnois}. Notice, in the case of symmetric inputs $I_L = I_R = I$, we can fit this relationship to the exponential
\begin{align}
\langle T \rangle \approx  A \exp [ B ( 1 - I) ].  \label{mdtcomps}
\end{align}
To understand why this is so, we can study the energy function associated with the system (\ref{compsimp}). Notice, this network is essentially the classic two neuron flip-flop Hopfield network. As has been shown before \cite{hopfield84}, for symmetric inputs the energy function for this network can be defined
\begin{align}
E[ u_R , u_L ] = H(I - u_R) H(I-u_L) - I \left[ H(I - u_R) + H(I - u_L ) \right],  \label{compseng}
\end{align}
so we can compute the energy difference between the winner-take-all and fusion states
\begin{align}
E[1,0] = - I, \ \ \ \ \ \ \  E[1,1] = 1 - 2I, 
\end{align}
respectively. By taking the difference between these two quantities, we find $\Delta E = 1 - I$, which well approximates the exponential dependence of the dominance times $T$, as shown by the fit in Fig. \ref{sccmpnois}{\bf (B)}. This provides the intuition as to this relationship.

In the same way, we can write down the energy function in the case where the inputs are non-symmetric $I_R \neq I_L$, which is \cite{chen01}
\begin{align}
E[u_R, u_L ] = H(I_L - u_R) H(I_R - u_L) - I_L H(I_L - u_R) - I_R H(I_R - u_L ), 
\end{align}
which means the energy depth of the right and left winner-take-all states are $\Delta E_R = 1- I_L$ and $\Delta E_L = 1- I_R$, respectively. Thus, as we observe in Fig. \ref{sccmpnois}{\bf (C)}, we expect the dominance time of each population to depend upon the strength of the other stimulus according to
\begin{align}
\langle T_R \rangle \approx A_R \exp \left[ B_R ( 1 - I_L ) \right], \ \ \ \ \ \ \langle T_L \rangle \approx A_L \exp \left[ B_L ( 1 - I_R ) \right]. 
\end{align}
Interestingly, this simple model agrees well with the qualitative predictions of Levelt propositions (i-iv) in this high contrast input regime. Now, we will see how including synaptic depression in the model generates distributions of dominance times that are more similar to those observed experimentally \cite{fox67,lehky95,brascamp06}.

\begin{figure}
\begin{center} \includegraphics[width=13cm]{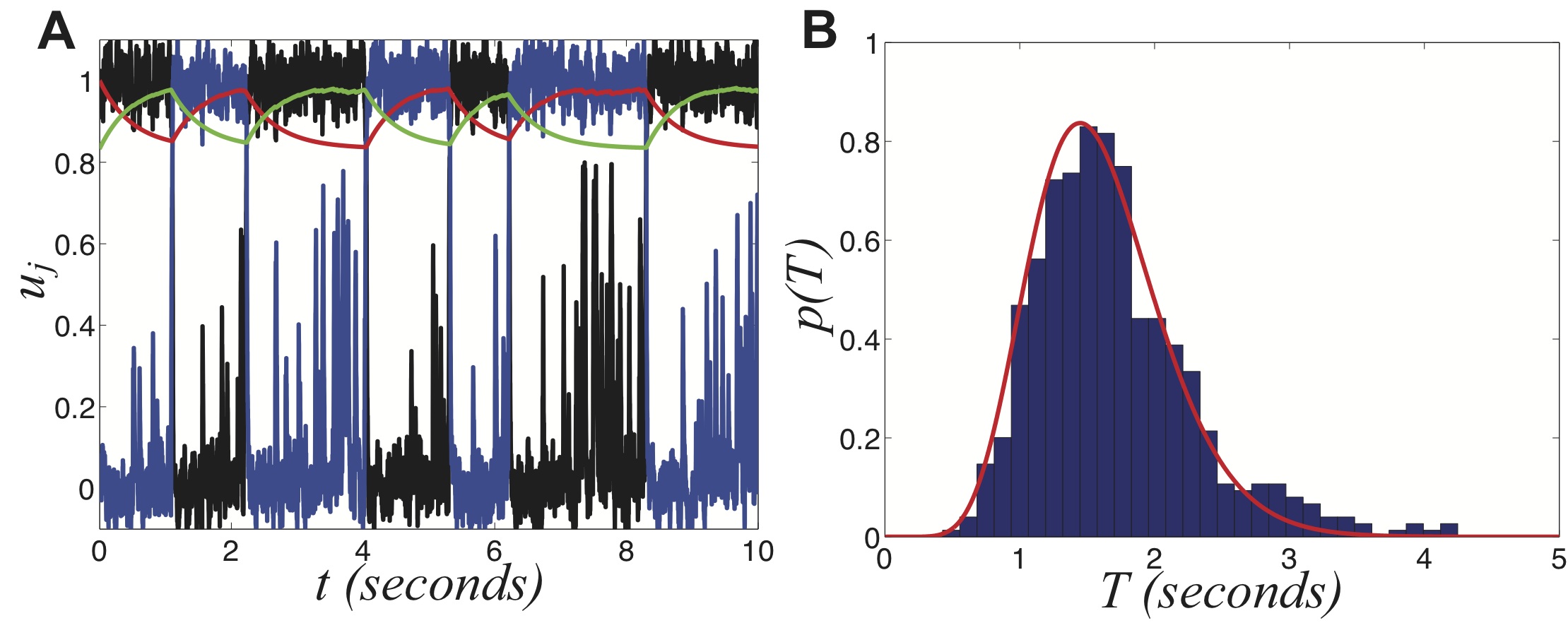} \end{center}
\caption{{\bf (A)} Single realization of the network (\ref{fulcompnet}) with depression and noise. Activity variables $u_R$ (black) and $u_L$ (blue) stay close to attractors at 0 and 1, aside from depression or noise induced switching. Depression variables $q_R$ (red) and $q_L$ (green) slowly exponentially change in response to the states of $u_R$ and $u_L$. {\bf (B)} Probability density $p(T)$ of dominance times $T$ sampled over 1000s, well fit by a gamma distribution (\ref{gamdist}). Parameters are $\ve = 0.036$, $\beta = 0.2$, $\tau = 50$, and $I = 0.8$.}
\label{compgam}
\end{figure}

\begin{figure}
\begin{center} \includegraphics[width=13cm]{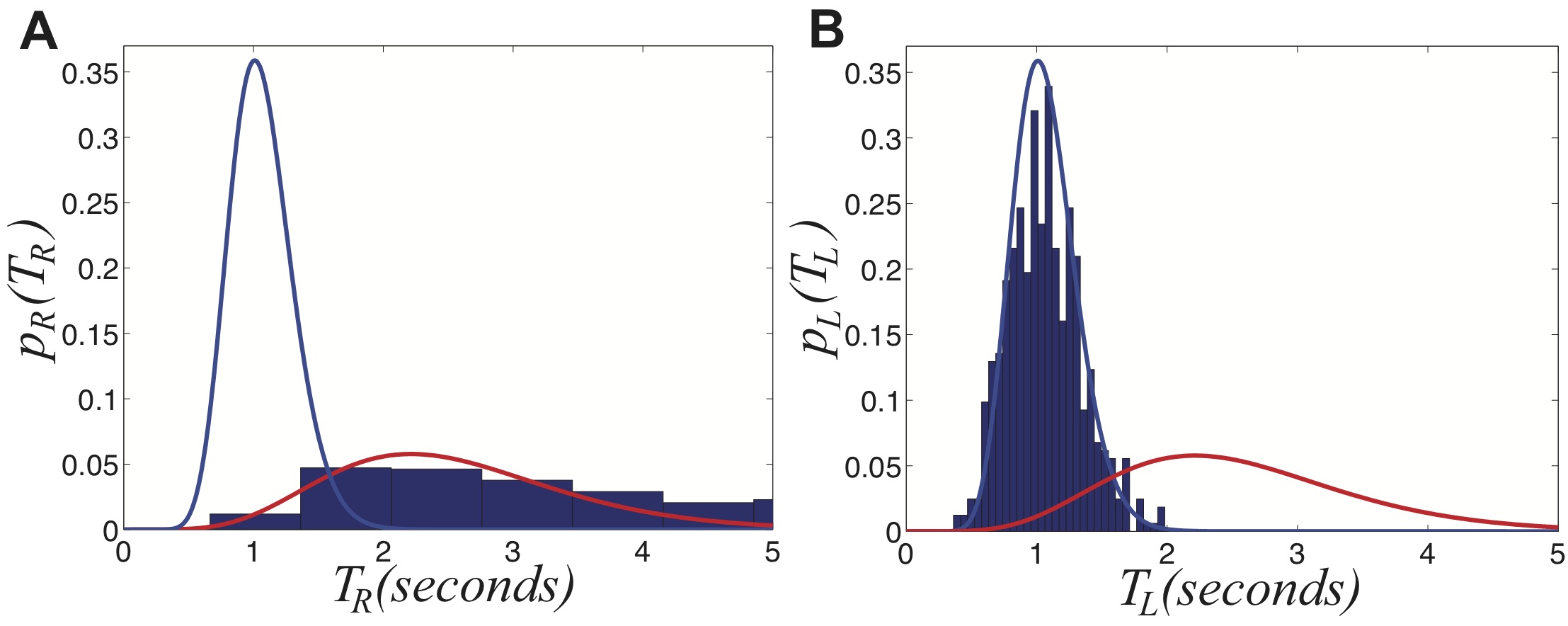} \end{center}
\caption{Distribution of dominance times for {\bf (A)} right and {\bf (B)} left populations fit with red and blue gamma distributions (\ref{gamdist}) respectively, in the network (\ref{fulcompnet}) with depression and noise in the case of asymmetric inputs $I_R = 0.82$ and $I_L=0.78$, sampled over 1000s. The right population has longer dominance times. Other parameters are $\beta = 0.2$, $\tau = 50$, and $\ve = 0.036$.}
\label{cgamasym}
\end{figure}

Finally, we show that the network with depression and noise generates gamma distributed dominance times, as the spatially extended system does. In addition, we provide some analytic intuition as to how gamma distributed dominance times may arise in the fast slow system. First, we display as single realization of the network (\ref{fulcompnet}) in Fig. \ref{compgam}{\bf (A)} along with a plot of an adiabatically computed energy function $E[u_R,u_L,q_R,q_L]$ for the system. To compute the energy function, we first note that in the limit of slow depression recovery time $\tau \gg 1$, we can assume the energy of the system will be defined simply by (\ref{compseng}) augmented by the synaptic scalings imposed by $q_R$ and $q_L$ \cite{mejias10}. In the fully general case, where inputs $I_R$ and $I_L$ may be asymmetric we have
\begin{align}
E[u_R,u_L,q_R,q_L] =& H(I_L - q_R u_R) H(I_R - q_L u_L) \nonumber \\ & - \frac{I_L}{q_R} H(I_L - q_R u_R) - \frac{I_R}{q_L} H(I_R - q_L u_L ).
\end{align}
A similar energy function was previously used in a model with spike frequency adaptation \cite{moreno07}. Here, we are able to derive the energy function from the model (\ref{fulcompnet}). Therefore, the energy gap between a winner-take-all state and the fusion state will be time-dependent, varying as the synaptic scaling variables $q_R$ and $q_L$ change. The energy difference between the right dominant state and fusion is
\begin{align}
\Delta E_R(t) = 1 - \frac{I_L}{q_R(t)}, \ \ \ \ \ \  \Delta E_L(t) = 1 - \frac{I_R}{q_L(t)},
\end{align}
for the right and left population respectively. 

Notice that dominance times of stochastic switching (Fig. \ref{compgam}) in (\ref{fulcompnet}) are distributed roughly according to a gamma distribution (\ref{gamdist}). Superimposing the probability density of right (left) dominance times on the left (right) probability density, we see they are reasonably separated. Using the analysis we performed for the spatially extended system, we could also show that depression improves discernment of the input contrast difference. Mainly here, we wanted to provide a justification as to the relationship between input strength and mean dominance times. Using energy arguments, we have provided reasoning behind why Levelt's propositions are still preserved in this model, when noise is included, even when switches are noise-induced. Increasing one input leads to a reduction in the energy barrier between the {\em other} population's winner-take-all state and the fusion state. This leads to the {\em other} population's dwell time being shorter.

\subsection{Switching between three percepts}

\begin{figure}
\begin{center} \includegraphics[width=13cm]{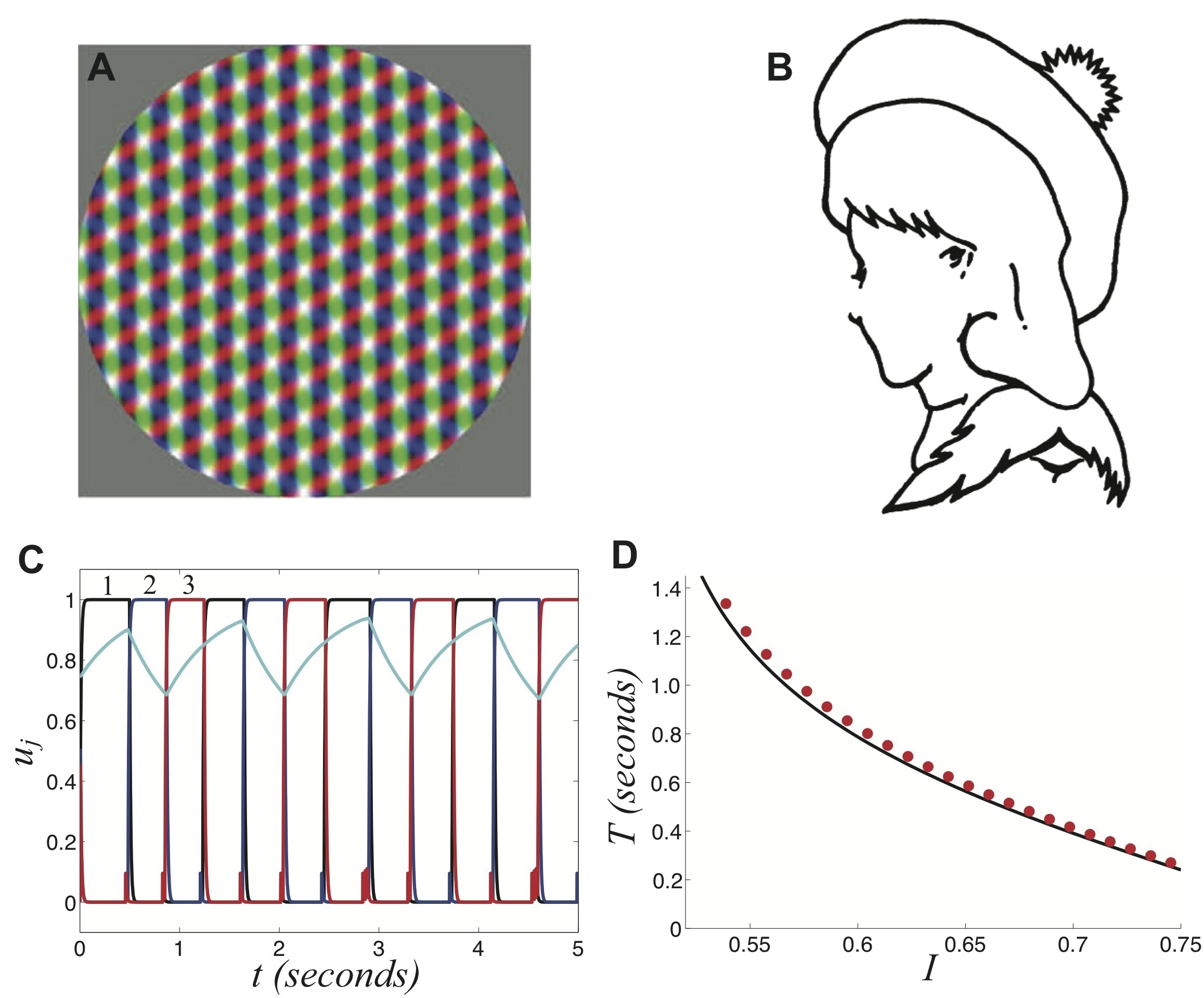} \end{center}
\caption{{\bf Perceptual tristability.} Examples of images with three possible interpretations. {\bf (A)} Three overlapping gratings. Redrawn with permission from \cite{naber10}. {\bf (B)} `Mother, father, and daughter.' Redrawn with permission from \cite{fisher68}. Staring at tristable images for long enough leads to the perception switching between the three possible interpretations. {\bf (C)} Numerical simulation of showing the activity variables $u_1$, $u_2$, $u_3$ and the second synaptic scaling variable $q_2$ (cyan) of the three population network (\ref{tresnet}) driven by symmetric stimulus $I = 0.7$. {\bf (D)} Relationship between the strength of the stimulus $I$ and the dominance times $T$ computed using fast-slow analysis (black) and numerics (red dots). Other parameters are $\beta = 1$ and $\tau = 50$.}
\label{tresperc}
\end{figure}

Finally, we will compare the transfer of information in competitive networks that process more than two inputs. Recently, experiments have revealed that perceptual multistability can switch between three or four different percepts \cite{fisher68,burton02,naber10,hupe12}. In particular, the work of \cite{naber10} characterized some of the switching statistics during the oscillations of perceptual tristability. Fig. \ref{tresperc}{\bf (A,B)} shows examples of tristable percepts. Since dominance times are gamma distributed and there is memory evident in the ordering of percepts, the process is also likely governed by some slow adaptive process in addition to fluctuations.

We will pursue the study of perceptual tristability in a competitive neural network model with depression and noise. In the case of three different percepts, a Heaviside firing rate (\ref{H}), and symmetric inputs $I_1 = I_2 = I_3 = I$, we study the system
\begin{subequations} \label{tresnet}
\begin{align}
\dot{u}_1 &= - u_1 + H(I - q_2 u_2 - q_3 u_3), \\
\dot{u}_2 &= -u_2 + H(I - q_1 u_1 - q_3 u_3), \\
\dot{u}_3 &= -u_3 + H(I - q_1 u_1 - q_2 u_2), \\
\tau \dot{q}_j &= 1 - q_j - \beta u_j q_j, \ \ \  \  j = 1,2,3.
\end{align}
\end{subequations}
We are interested in rivalrous oscillations, which do arise in this network for certain parameter regimes (Fig. \ref{tresperc}{\bf (C)}). As per our previous analysis, we perform a fast-slow decomposition of our system. In the case of symmetric inputs, we use our techniques to compute the dominance time $T$ of a population as it depends on input strength $I$. Our analysis follows along similar lines to that carried out for the two population network, where we assume $\tau \gg 1$. We find
\begin{align}
T = \tau \ln \left[ \frac{(1-I)(1+ \beta) + \sqrt{(1+ \beta)(1-I)[3I(1+ \beta) + \beta - 3]}}{2[(1+ \beta) I - 1]} \right],
\end{align}
which compares very well with numerically computed dominance times in Fig. \ref{tresperc}{\bf (D)}. While perceptual tristability has not been explored very much experimentally \cite{fisher68,naber10,hupe12}, observations that have been made suggest that relationships between mean dominance time and input contrast may be similar to the two percept case \cite{hupe12}. In our model, we see that as the input strength is increased, dominance times decrease. One other important point is that percept dominance occurs in the same order every time (Fig. \ref{tresperc}{\bf (C)}): one, two, three. There are no ``switchbacks." We will show that this can occur in the noisy regime, which degrades information transfer.

\begin{figure}
\begin{center}  \includegraphics[width=13cm]{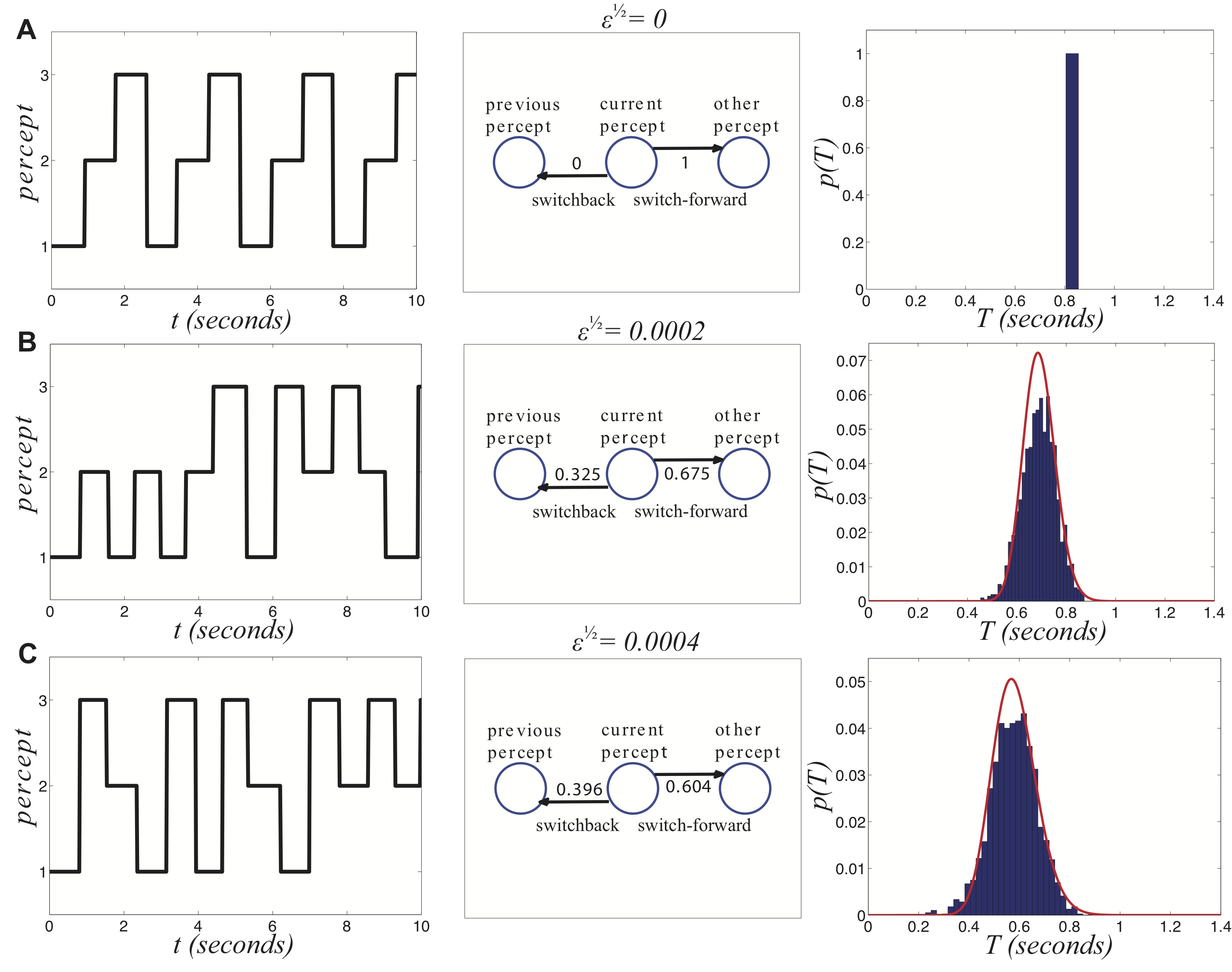}  \end{center}
\caption{Noise degrades two sources of information provided by dominance switches. {\bf (A)} In the absence of noise, switches always move ``forward," so that the previous percept perfectly predicts the subsequent percept. Dominance times accumulate at a single value too. {\bf (B)} For slightly higher levels of noise ($\sqrt{\ve} = 0.0002$), ``switch backs" can occur where the subsequent percept is the same as the previous percept. Also, the distribution of dominance times spreads. {\bf (C)} For stronger noise ($\sqrt{\ve} = 0.0004$). Other parameters are $I_0 = 0.6$, $\beta = 1$, and $\tau = 50$. }
\label{tresdepno}
\end{figure}

\begin{figure}
\begin{center} \includegraphics[width=6cm]{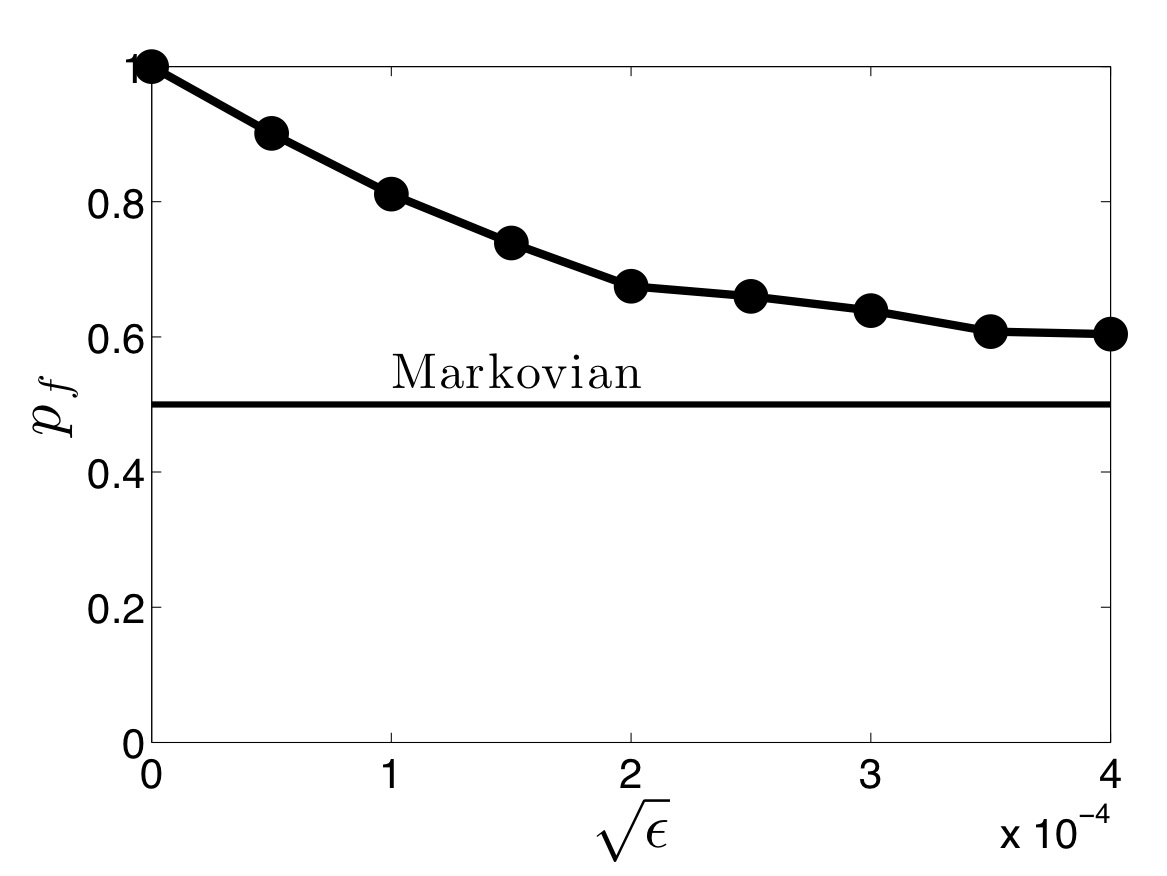} \end{center}
\caption{The probability of a switch being in the forward direction in simulations of (\ref{tresdn}) as a function of the amplitude $\ve$ of noise. As $\ve$ increases, network switches behave in more of a Markovian way, not reflecting any memory of the previous percept. Therefore, information of the previous percept is lost as soon as a switch occurs.}
\label{markov}
\end{figure}

Now, we seek to understand how noise alters the switching behavior when added to the deterministic network (\ref{tresnet}). Thus, we discuss the three population competitive network with noisy depression
\begin{subequations} \label{tresdn}
\begin{align}
\dot{u}_1 &= - u_1 + H(I - q_2 u_2 - q_3 u_3), \\
\dot{u}_2 &= -u_2 + H(I - q_1 u_1 - q_3 u_3), \\
\dot{u}_3 &= -u_3 + H(I - q_1 u_1 - q_2 u_2) , \\
\tau \dot{q}_j &= 1 - q_j - \beta u_j q_j + \xi_j, \ \ \  \  j = 1,2,3,
\end{align}
\end{subequations}
where $\xi_j$ are identical independent white noise processes with variance $\ve$. In Fig. \ref{tresdepno}, we show the noise in (\ref{tresdn}) degrades two pieces of information carried by dominance switches: the switching time and the direction of switching. Notice that as the amplitude of noise $\ve$ is increased, the dominance times become more spread out. Thus, there is a less precise characterization of the input strength in the network. Concerning the direction of switching, we see that the introduction of noise makes ``switch backs" more likely. We define a ``switch back" as a series of three percepts that contains the same percept twice (e.g. $1 \to 3 \to 1$). This as opposed to a ``switch forward," which contains all three percepts (e.g. $1 \to 3 \to 2$). Statistics like these were analyzed from psychophysical experiments of perceptual tristability, using an image like Fig. \ref{tresperc}{\bf (A)} \cite{naber10}. The main finding of \cite{naber10} concerning this property is that switch forwards occurred more often than chance would suggest. Therefore, they proposed that some slow process may be providing a memory of the previous image. We suggest short term depression as a candidate substrate for this memory. As seen in Fig. \ref{tresdepno}, the bias in favor of switching forward persists even for substantial levels of noise. The idea of short term plasticity as a substrate of working memory was also recently proposed in \cite{mongillo08}. Our results extends this idea, suggesting synaptic mechanisms of working memory may be useful in visual perception tasks, such as understanding ambiguous images. In Fig. \ref{markov}, we show that the process of dominance switching becomes more Markovian as the level of noise $\sqrt{\ve}$ is increased even a modest amount. In the limit of large noise, the likelihoods of ``switch forwards" and ``switch backs" are the same.

\section{Discussion}

Mechanisms underlying stochastic switching in perceptual rivalry have been explored in a variety of psychophysical \cite{fox67,lehky95,brascamp06}, physiological \cite{leopold96,blake02}, and theoretical studies \cite{matsuoka84,laing02,moreno07}. Since psychophysical data is widely accessible, it can be valuable to use the hallmarks of its statistics as benchmarks for theoretical models. For instance, the fact that dominance time distributions are unimodal functions peaked away from zero suggests that some adaptive process must underlie switching in addition to noise \cite{laing02,brascamp06,shpiro09}. In addition, \cite{moreno11} recently suggested the visual system may sample the posterior distribution of interpretations of bistable images. This type of sampling can be well modeled by attractor networks analogous to those presented here \cite{moreno07}. Therefore, many dominance time statistics from perceptual rivalry experiments can be employed as points of reference for physiologically based models of visual perception. New data now exists concerning tristable images showing this process also likely is guided by a slow adaptive process in addition to fluctuations \cite{naber10}.

We have studied various aspects of competitive neuronal network models of perceptual multistability that include short term synaptic depression. First, we were able to analyze the onset of rivalrous oscillations in a ring model with synaptic depression \cite{york09,kilpatrick10b}. Stimulating the network with a bimodal input leads to winner-take-all solutions, in the form of single bumps, in the absence of synaptic depression. As the strength of synaptic depression is increased, the network undergoes a bifurcation which leads to slow oscillations whose timescale is set by that of synaptic depression. Each stimulus peak is represented in the network by a bump whose dominance time is set by the height of each peak. Thus, synaptic depression reveals information about the stimulus that would otherwise be masked by the lateral inhibitory connectivity of the network. The inclusion of noise in the network leads to dominance times that are exponentially (gamma) distributed in the absence (presence) of synaptic depression. Motivated by recent work exploring how visual perception may exploit Bayesian sampling on posterior distributions \cite{kersten04,hohwy08,moreno11}, we considered the simple task of an observer trying to infer the contrast of stimuli based on dominance times. We found Bayesian sampling of the dominance times discerning input contrast differences better as switches become more depression driven and less noise-driven. Thus, short term depression improves information transfer of networks that process ambiguous images in multiple ways.

We also used energy methods in simple space-clamped neural network models to understand how a combination of noise and depression interact to produce switching in competitive neural networks. Using the energy function derived by Hopfield for analog neural networks, we justify the exponential dependence of dominance times upon input strength in purely noise-driven switching. Studying an adiabatically derived energy function for the case of slow depression, we also show how depression works to reduce the energy barrier between winner-take-all states, leading to the slow timescale that defines the peak in depression-noise generated switches. Finally, using a three population space-clamped neural network, we analyzed depression and noise generated switching that may underlie perceptual tristability. We found this network also sustained some of the same relationships between input contrast and dominance times as the two population network. Also, we found that when switches are generated by depression there is an ordering to the population dominance that is lost when switches are noise generated. This is due to the memory generated by short term depression \cite{mongillo08}, so the switching process is non-Markovian due to the inherent slow timescale in the background. However, even small amounts of noise can wash this memory away. Thus, since recent psychophysical experiments reveal a non-Markovian property to percept ordering, this provides further support for the idea that a slow adaptive process underlies percept switching.

Note to analytically study the relationship between dominance times and input contrast in the noisy system, we resorted to a simple space-clamped neural network. In future work, we plan to develop energy methods for spatially extended systems like (\ref{unoise}). Such methods have seen success in analyzing stochastic partial differential equation models such as Ginzburg-Landau models \cite{e04}. Energy functions have recently been developed for neural field models, but have mostly been studied as a means of determining global stability in deterministic systems \cite{wu02,kubota05,coombes07}. We proposed that by deriving the specific potential energy of spatially extended neural fields, it may be possible to approximate the transition rates of solutions from the vicinity of one attractor to another. In the system (\ref{unoise}), there should be some separatrix between the two winner-take-all states that must be crossed in order for a transition to occur. The least action principle states that there is even a specific point on this separatrix through which the dynamics most likely flows \cite{e04}. Finding this with an energy function in hand would be straightforward would allow us to relate the parameters of the model to the distribution of dominance times. This would provide a better theoretical framework for interpreting data concerning rivalry of spatially extended images, such as those that produce waves \cite{wilson01}.

\bibliographystyle{siam}
\bibliography{switch}

\end{document}